\documentclass[12pt,showkeys,pra,groupedaddress,floatfix,nofootinbib,%
longbibliography]{revtex4-2}

\usepackage{graphicx}
\usepackage{color}
\usepackage{amsfonts}
\usepackage{eurosym}

\usepackage{epstopdf}
\usepackage{xspace}

\renewcommand {\d} {{\rm d}}

\newcommand {\E}  {{\varepsilon}}
\newcommand {\om}  {{\omega}}
\newcommand {\Om}  {{\Omega}}

\newcommand {\bfr} {{\bf r}}
\newcommand {\bfv} {{\bf v}}

\newcommand {\bfF} {{\mathbf F}}

\newcommand {\lamu}  {{\lambda_{\rm u}}}

\newcommand {\Nd}  {{N_{\rm d}}}

\newcommand{\MBNExplorer}{\textsc{MBN Explorer}\xspace}
\newcommand{\MBNStudio}{\textsc{MBN Studio}\xspace}

\begin{document}


\title{Narrowband $\gamma$-ray radiation generation by
acoustically driven crystalline undulators}

\author{Konstantinos Kaleris}
\email{kkaleris@hmu.gr}
\author{Evaggelos Kaselouris}
\author{Vasilios Dimitriou}
\author{Emmanouil Kaniolakis-Kaloudis}
\author{Makis Bakarezos}
\affiliation{Institute of Plasma Physics \& Lasers,
Hellenic Mediterranean University, Tria Monastiria, 74100 Rethymno, Greece}
\affiliation{Physical Acoustics \& Optoacoustics Laboratory,
Dept. of Music Technology \& Acoustics, Hellenic Mediterranean University,
74100 Rethymno, Greece}
\author{Michael Tatarakis}
\affiliation{Institute of Plasma Physics \& Lasers,
Hellenic Mediterranean University, Tria Monastiria, 74100 Rethymno, Greece}
\affiliation{Dept. of Electronics Engineering, Hellenic Mediterranean University, Romanou 3, Chalepa, 73133 Chania, Greece}
\author{Nektarios A. Papadogiannis}
\email{npapadogiannis@hmu.gr}
\affiliation{Institute of Plasma Physics \& Lasers,
Hellenic Mediterranean University, Tria Monastiria, 74100 Rethymno, Greece}
\affiliation{Physical Acoustics \& Optoacoustics Laboratory,
Dept. of Music Technology \& Acoustics, Hellenic Mediterranean University,
74100 Rethymno, Greece}
\author{Gennady B. Sushko}
\author{Andrei V. Korol}
\email{korol@mbnexplorer.com}
\author{Andrey V. Solov'yov}
\email{solovyov@mbnresearch.com}
\affiliation{MBN Research Center, Altenh\"{o}ferallee 3,
60438 Frankfurt am Main, Germany}


\date{\today}%

\begin{abstract}
In this paper we present a novel scheme for the controlled generation of
of tunable narrowband $\gamma$-ray radiation by
ultra-relativistic positron beams inside acoustically driven periodically
bent crystals.
A novel acoustic crystalline undulator is presented, in which
excitation of a silicon single crystal along the (100) planar direction
by a piezoelectric transducer periodically modulates
the crystal lattice in the [100] axial direction.
An ultra-relativistic positron beam is directed diagonally
into the crystal and
propagates along the (110) planes.
The lattice modulation forces the positrons to follow periodic trajectories,
resulting in the emission of undulator radiation in the MeV range.
A computational methodology for the design and development of such  acoustically
based light sources is presented together with the results of simulations
demonstrating the favourable properties  of the  proposed technology.
The longitudinal acoustic strains induced in the crystal by  high-frequency
piezoelectric elements are calculated by finite element simulations.
The resulting bending profiles of the deformed crystal planes are
used as geometrical conditions in  the relativistic molecular dynamics simulations
that calculate the positron  trajectories and the spectral distribution of the
emitted radiation.
The results show a strong enhancement of the emitted radiation within a narrow
spectral band defined by the bending period, demonstrating the feasibility and
potential of the proposed technology.

\end{abstract}

\maketitle

\section{Introduction  \label{Introduction}}

The controlled production of narrow-band gamma radiation in the MeV range and
beyond is one of the major challenges of modern physics, with significant
applications expected in a wide range of fields, including fundamental science,
industry, biology and medicine
\cite{AlbertThomas_PlasmaPhysContrFusion_v58_103001_2016,
Rehman_EtAl_ANE_v105_p150_2017,Kraemer_EtAl-ScieRep_v8_p139_2018,
NextGenerationGammaRayLS2022,CLS-book_2022}.

One of the novel approaches to the production of tunable intense gamma-ray
radiation is based on the use of Crystalline Undulators (CU)
\cite{ChannelingBook2014,CLS-book_2022}.
In a CU the radiation is emitted by ultrarelativistic leptons
(electrons or positrons) that channel in a
Periodically Bent Crystal (PBC).
As a result, radiation is emitted by two main mechanisms:
(i) the channeling radiation (ChR) due to the channeling oscillations
\cite{ChRad:Kumakhov1976},
(ii) the CU radiation (CUR) due to the periodicity of the trajectory of a
particle following the bending profile.
Typically, the spatial period $\lambda_{\rm u}$ of the bending is much
larger than that of the channeling oscillations.
Therefore, the characteristic frequencies $\om$ (harmonics) of ChR and CUR are well separated, satisfying the condition
$\om_{\rm CUR} \ll \om_{\rm ChR}$
\cite{KSG1998,KSG_review_1999}.

The feasibility of the CU concept has been verified theoretically by
analysing the essential conditions and limitations that must be met
\cite{EnLoss00,KSG_review_1999,KSG_review2004,ChannelingBook2014}.
Theoretical and experimental studies of the CU and CUR phenomena are
in the focus of the current Horizon Europe EIC-Pathfinder-2021 project
TECHNO-CLS \cite{TECHNO-CLS}.
This project aims at the practical realisation of novel intense gamma-ray
light sources which can be created by exposing oriented crystals of
different geometries (linear, bent, periodically bent) to ultrarelativistic
beams of electrons and positrons \cite{CLS-book_2022}.
The advantage of using Periodically Bent Crystals (PBC) is that by varying
the amplitude $a$ and the period $\lambda_{\rm u}$ of the bending
it is possible to optimize the characteristics of the emitted radiation for given
parameters of the incident beam.
However, the operational efficiency of such a device depends
strongly on the quality of the periodic bending \cite{ChannelingBook2014}.
One of the most challenging technological tasks is the production of
periodically bent crystals.

Several techniques have been developed to produce statically bent crystals.
Some of these include the surface deformation methods such as
mechanical grooving
\cite{Bellucci_EtAl-PRL_v90_034801_2003,Bagli-EtAl_EPJC_v74_p3114_2014},
laser ablation \cite{Balling_EtAl-NIMB_v267_p2952_2009},
strip deposition
\cite{Guidi_EtAl-NIMB_v234_p40_2005,Guidi_EtAl-APL_v90_114107_2007,%
Guidi_EtAl-ThinSolFilms_v520_p1074_2011},
ion implantation \cite{Bellucci_EtAl:APL_v107_064102_2015},
laser pulse melting \cite{Carraro_EtAl:ApplSurcScie_v509_145229_2020},
and
sandblasting \cite{Camattari-EtAl:JApplCryst_v50_p145_2017}.
The stress generated at the crystal surface propagates into the bulk,
resulting in the bending of the crystal planes.
At present, surface deformation methods can achieve bending periods of several hundred microns.
To reduce the period one can rely on the production of graded
composition strained layers in a silicon-germanium superlattice
\cite{Breese:NIMB_v132_p540_1997,MikkelsenUggerhoj:NIMB_v160_p435_2000,
AvakianEtAl-NIMA_v508_p496_2003}.
Replacement of a fraction of Si atoms by Ge atoms leads to the
bending of the crystallographic directions \cite{Dickers_EtAl:EPJD_v78_77_2024}.
A similar effect can be achieved by doping diamond with boron during
the synthesis process
\cite{ThuNhiTranThi_JApplCryst_v50_p561_2017,BackeEtAl:arXiv_2404.15376}.
The advantage of a diamond crystal is its radiation hardness, which
allows the lattice integrity to be maintained in the environment of very
intense beams \cite{Uggerhoj_RPM2005}.

Dynamic bending in crystals can be achieved by the propagation of an
acoustic wave (AW) along a particular crystallographic direction
\cite{KaplinPlotnikovVorobev1980,BaryshevskyDubovskayaGrubich1980,
IkeziLinLiuOhkawa1984,Mkrtchyan_EtAl:PLA_v126_p528_1988,
Dedkov:PhusStatSol-b_v184_p535_1994,
KSG1998,KSG_review_1999}.
As mentioned in Ref. \cite{KSG_review_1999}, this can be achieved
by attaching a piezoelectric element to the crystal and generating
radio frequencies.
The advantage of this method is its flexibility: the bending
amplitude and period can be changed by varying the AW intensity and
frequency \cite{EnLoss00}.
Within the TECHNO-CLS project \cite{TECHNO-CLS}, experimental approaches
for inducing AWs to generate PBCs are being explored for the first time and
evaluated for their effectiveness in practical implementation and use
in channeling experiments.
An alternative method to induce acoustic waves  in crystals using ultrafast
pulsed laser excitation has recently been described in
Ref. \cite{KalerisEtAl:ApplPhys_v129_p527_2023}, but is
beyond the scope of this paper.

In this paper we present a roadmap for the development of
Acoustically driven CUs (A-CUs) and their use for the
production of narrowband gamma radiation under real experimental
conditions.
The roadmap includes the introduction of a novel CU
implementation based on 
acoustic excitation of a Si crystal via a multi-MHz piezoelectric transducer, and computational simulations of
(i) the formation of travelling AWs inside the crystal for
periodic spatial modulation of its planes,
(ii) the channeling of ultra-relativistic positrons
and the resulting light emission.
To the best of our knowledge, no computational study of gamma-ray production via A-CUs has been performed to date.
The aim of this work is to demonstrate the feasibility of using
A-CUs for the controlled generation of strong narrowband
gamma rays for the first time at the proof-of-principle level.
It also provides a theoretical and computational framework for
the parametric study and design of A-CU devices, which is
essential for the development and future experimental evaluation of
of A-CU prototypes.
In parallel to the described computational analysis,
the authors are working towards the realisation of the first experiments
with A-CUs, which are planned for the near
near future.

The paper is organized as follows.
Section \ref{Methods_A-CU} describes the operating principles of the
A-CU scheme. Section \ref{Feasibility} discusses the conditions that must be
fulfilled to consider A-CU as a feasible scheme.
Details on the simulations of structural modulation of the crystalline
lattice due to the AW propagation are given in Section \ref{Methods_FEM}.
The general methodology for the simulation of the passage of ultra-relativistic
particles in crystalline environments and related phenomena is described in
Section \ref{MBN_Methodology}.
The results of the numerical simulations are presented and discussed in
Section \ref{Results}.
The exemplary case study considered is related to the 20 GeV positrons
channeling along the (110) planar direction in acoustically excited
silicon crystals.
Section \ref{Conclusions} summarises the conclusions of this work
and presents future perspectives.

\section{Methods \label{Methods}}

\subsection{Acoustically driven crystalline undulators (A-CUs)
\label{Methods_A-CU}}

\begin{figure}
    \centering
    \includegraphics[scale=1.0]{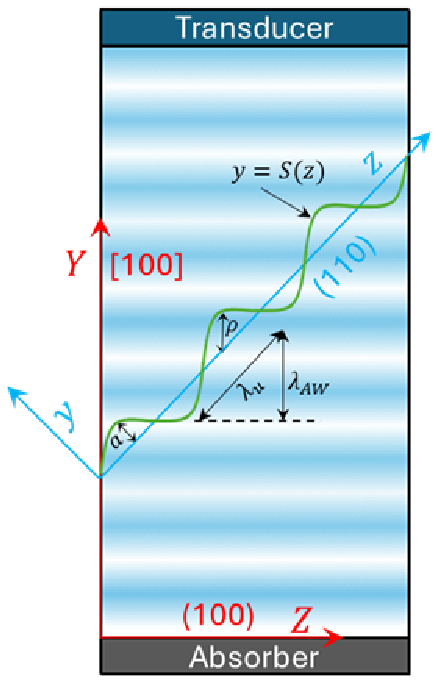}
    \caption{\textcolor{black}{Schematic diagram of an A-CU device for periodic bending of the crystal lattice. } A plane longitudinal AW excited by the transducer along the  $[100]$ crystallographic direction causes a periodic transverse deformation $y=S(z)$ of the $(110)$ planes in the crystal. The absorber dissipates the propagating acoustic energy. The gradient shading illustrates the periodic deformation (compression and rarefaction) of the crystal structure caused by the AW.
   %
    }
    \label{Figure:A-CU-scheme}
\end{figure}

\textcolor{black}{In this section, the operation principles of the novel A-CU scheme are presented. The schematic diagram of such a device is shown in Fig. \ref{Figure:A-CU-scheme}. At the heart of the device lies a single crystal with a cubic unit cell and near zero dislocation density, for example high quality Si or Ge. A piezoelectric transducer attached to one end of a crystal induces harmonic longitudinal acoustic waves which can be either standing or traveling, depending on the design. In the latter case, which is of interest here, an absorbing or non-reflecting structure is used on the other end of the crystal to dissipate the propagating acoustic energy.}
The longitudinal AW of wavelength $\lambda_{\rm AW}$ excited by the
transducer produces periodic regions of compression and rarefaction
deformation in the bulk of the crystal (shown by the gradient shading)
along the $[100]$ crystallographic direction. The deformation, in turn, leads to the periodic bending of the $(110)$
planes, which is characterised by the bending profile $y=S(z)$, where
the $z$-axis is along the plane and the $y$-axis is along $[110]$.
The angle between the $[100]$ and $[110]$ axes is $45^{\circ}$,
so the bending period $\lambda_{\rm u}$ and $\lambda_{\rm AW}$ are
related according to $\lambda_{\rm u}=\sqrt{2}\lambda_{\rm AW}$.
The amplitude $a$ of the periodic bending is directly related to the
amplitude $A_{\rm AW}$ of the acoustic wave or, equivalently,
to the displacement $\rho$ of the $(100)$ planes from their original position. \textcolor{black}{It should be noted here that layouts as the one presented in Fig. \ref{Figure:A-CU-scheme} are known in the scientific community as Acousto-Optic Modulators (AOMs). Such devices are commonly used to control the energy of laser beams by exploitation Bragg diffraction due to the periodic modulation of the refractive index induced by the acoustic wave. For this reason, the proposed A-CU scheme will be alternatively mentioned as AOM-type A-CU.}

A beam of ultra-relativistic positrons (not shown in the figure)
is incident along the $(110)$ planar direction.
The undulator motion of the particles passing through the crystal following
the periodic bending results in the emission of the intense CUR,
which can be well within the gamma-ray regime under conditions of
sufficiently small period $\lambda_{\rm u}$, sufficiently large
amplitude $a$, crystal thickness $L$ and beam energy $\E$.
An analysis of the constraints on these parameters specific to the A-CU
scheme is given in Section \ref{Feasibility}.

\begin{figure}
    \centering
    \includegraphics[width=1\linewidth]{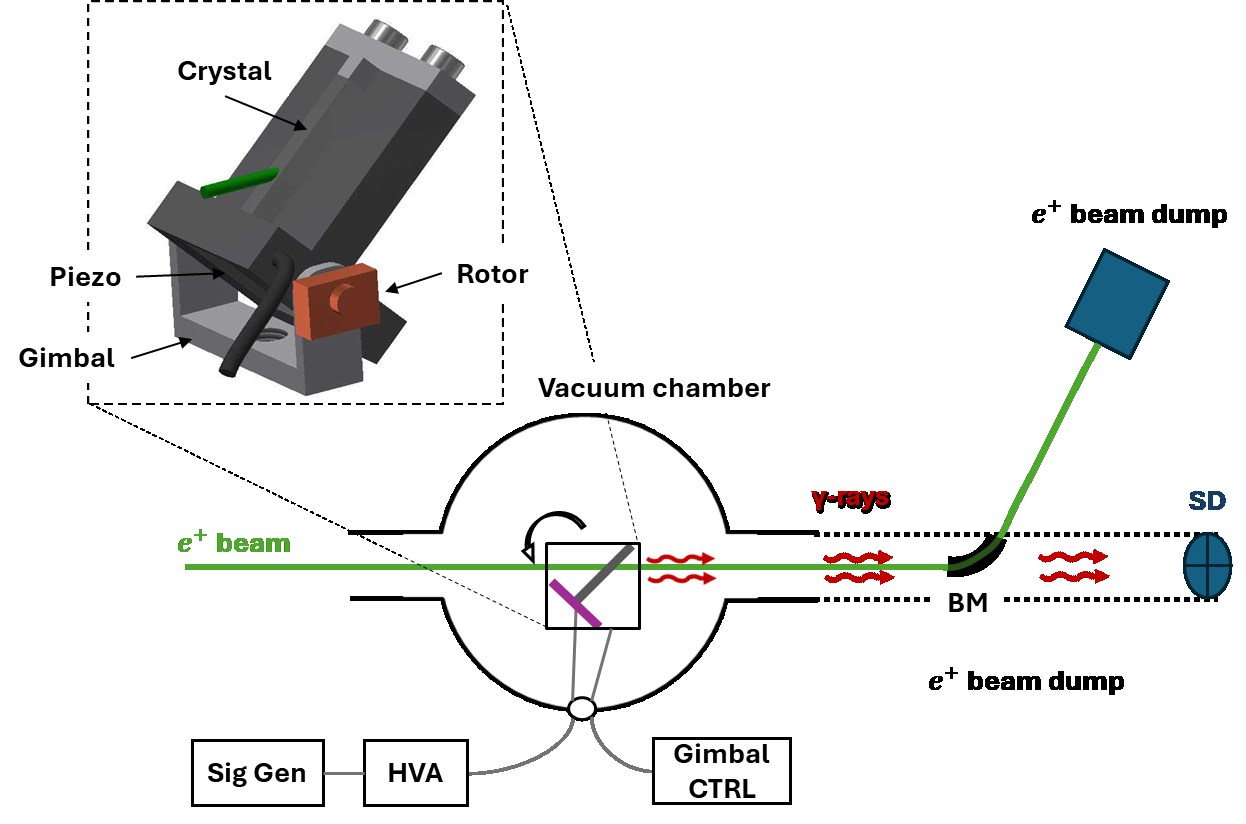}
    \caption{Schematic diagram of the experimental setup for the generation of $\gamma$-ray radiation using an AOM-type acoustically-driven crystalline undulator.
    }
    \label{Figure:setup}
\end{figure}

The proposed scheme represents an experimentally feasible technology for the production of gamma-ray radiation generation by means of oriented crystals as it can lead to relatively simple and manageable A-CU devices and setups.
A complete experimental setup for gamma-ray radiation generation
based on the simulated prototype AOM-type A-CU device is presented in Fig.
\ref{Figure:setup}.
The core of the device is a monocrystal acoustically driven by a
piezoelectric transducer with a resonant frequency in the order of
tens of MHz.
Such piezoelectric elements are usually made by coating thin
crystal plates with high piezoelectric constants, e.g. ZnO or
SiO$_2$, with conducting metal films on the two surfaces.
The piezoelectric element is driven by a tunable multi-MHz
high-voltage amplifier (HVA) capable of delivering 100 V or more
to the transducer, causing a harmonic surface displacement of
the piezoelectric element of several nanometers.
The HVA is controlled by a tunable radio frequency (RF) signal
generator, delivering low power electrical signals in the multi-MHz
range.
The piezoelectric transducer is bonded with the Si or Ge undulation
monocrystal, inducing multi-MHz acoustic waves in the crystal with
wavelengths of a few to several hundreds of $\mu$m.
A suitable housing is used to keep the crystal stable on the
piezoelectric transducer and to avoid unwanted acoustic reflections
from the boundaries.
In addition, the crystal is cut diagonally at the free end and
covered with a sound-absorbing material, such as rubber, to ensure
the absence of standing waves in the bulk of the crystal,
which would have a
detrimental effect on the efficiency of the
radiation generation.
The housing is mounted on a high-precision vacuum compatible motorized
gimbal mechanism that allows for accurate alignment of the (110)
planar direction of
the undulation crystal with the positron beam.
The entire instrument is placed in a vacuum chamber. This prevents
the measurement signals due to the ionisation of the ambient air by by the positron beam and the secondary gamma rays produced.
The chamber has the appropriate feed-through connections for the
electrical drive and suitable windows for the entry and exit of the
positron beam and the emitted radiation.
A bending magnet (BM) is used to deflect the positrons, separating their
trajectories from those of the gamma-rays.
The positron beam is terminated on a beam dump while the radiation is
collected by a scintillator detector (SD), e.g. NaI
\cite{Bandiera_EtAl:EPJC_v81_284_2021,BackeEtAl:NIMB_266_p3835_2008}.

\subsection{Feasibility of an A-CU  \label{Feasibility}}

In this section we discuss the conditions that  must be met in order
to consider an A-CU as a feasible scheme.
In carrying out this analysis, particular attention will be paid to
to the specific constraints that must be imposed on the dynamic bending of a crystalline medium by the propagation of an AW.

As a case study, we consider a crystal of thickness $L$, measured
along the $(100)$ planar direction
(the $Z$ axis, see  Fig. \ref{Figure:A-CU-scheme}), in which
a plane longitudinal AW of frequency $\nu$ is excited along the
the $\langle 100 \rangle$ crystallographic
axis (the $Y$ axis).
The AW results in the deformation of the crystal lattice.
The instantaneous displacement $\rho$ of the lattice element from its
position in the non-deformed crystal can be written as
$\rho(t)
=
A_{\rm AW}\cos\left(k_{\rm AW}Y + 2\pi\nu t\right)$ where
$A_{\rm AW}$ and $k_{\rm AW}=2\pi/\lambda_{\rm AW}$
are the AW amplitude and wave number, and
$\lambda_{\rm AW}$ stands for the AW wavelength.

To derive a profile of the (ideal) harmonic bending of the $(110)$ planes
let us consider a $(110)$ plane that crosses the $Y$ axis at $Y=Y_0$.
Let $(y,z)$ be a Cartesian coordinate system with its origin at
$(Y_0, Z=0)$.
The $y$ direction is aligned with the
$\langle 110 \rangle$ crystallographic axis, the $z$ axis is chosen along the non-deformed
$(110)$ plane.
Then, assuming the long-wavelength limit for the AW, i.e. $k_{\rm AW}A_{\rm AW}\ll 1$,
one derives
\begin{eqnarray}
\displaystyle
y(z;t)
\approx
a \cos\left({2\pi z \over \lamu} + \phi_0\right)\,,
\label{Geometry:eq.07b}
\end{eqnarray}
where $a = A_{\rm AW}/\sqrt{2}$, $\lamu = \lambda_{\rm AW} \sqrt{2}$
and the phase factor $\phi_0 = k_{\rm AW}Y_0 + 2\pi\nu t$.

\subsubsection{General conditions for the CU feasibility
\label{Restrictions_04}}

The conditions which must be met in order to treat any crystalline undulator
 as a feasible scheme have been formulated some time ago
\cite{KSG1998,KSG_review_1999,KorolSolovyov:EPJD_CLS_2020,CLS-book_2022}:
\begin{eqnarray}
\left\{
\begin{array}{ll}
C = 4 \pi^{2} \E a/U^{\prime}_{\max}
\lambda_{\mathrm{u}}^{2} \ll 1
&
\mbox{-- stable channeling,}
\\
d<a \ll \lambda_{\mathrm{u}}
&
\mbox{-- Large-Amplitude regime},
\\
N = L_{\rm u}/\lamu \gg 1
& \mbox{-- large number of periods},
\\
L_{\rm u} \lesssim \min\Bigl[L_{\mathrm d}(C),L_{\rm att}(\om)\Bigr]
& \mbox{  -- account for dechanneling and photon attenuation.}
\end{array}
\right.
\label{Restrictions_03:eq.01}
\end{eqnarray}
A short commentary on these conditions is presented below.

\begin{figure}[ht]
\centering
\includegraphics[scale=0.4,clip]{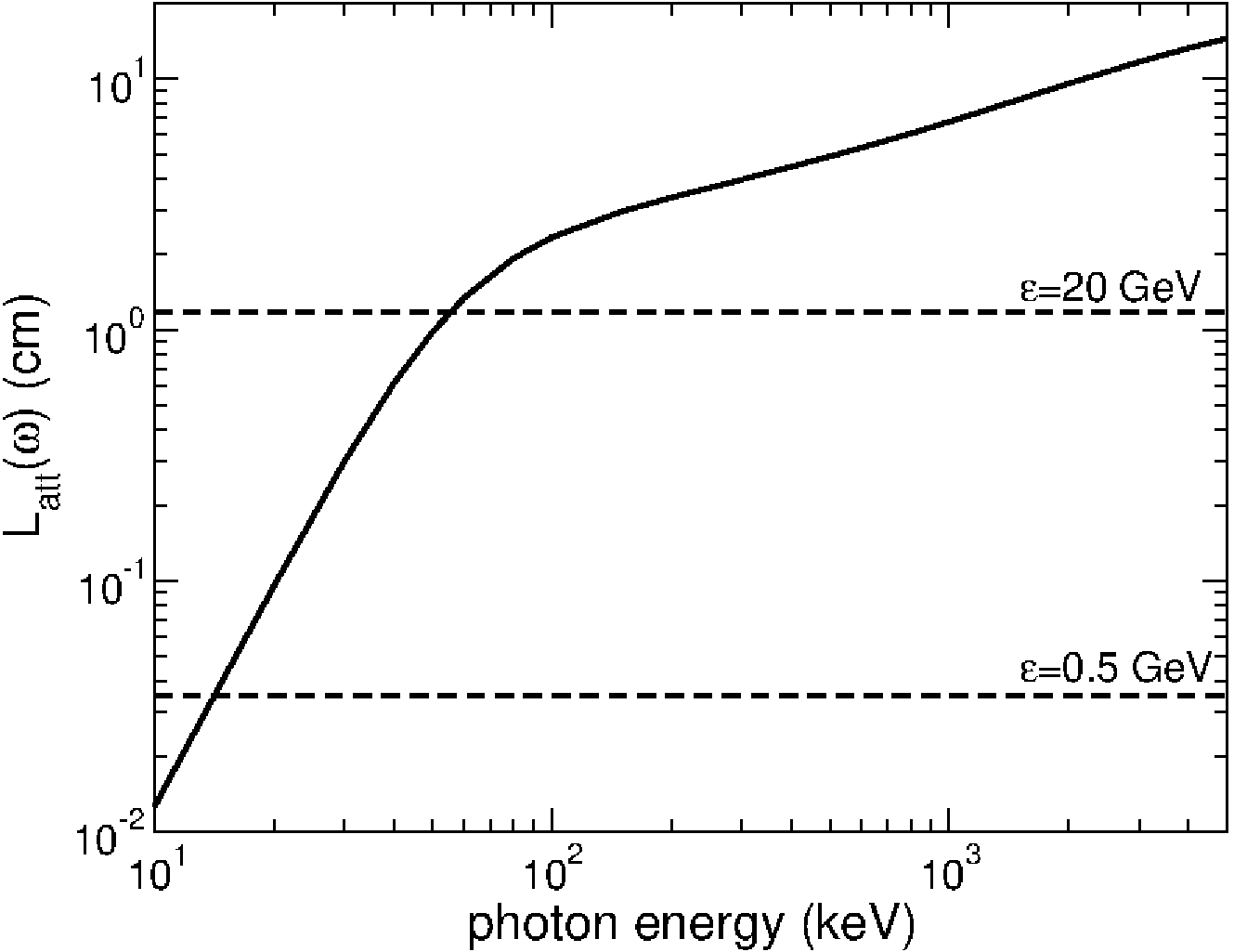}
\caption{
Photon attenuation length (solid line) versus photon energy for
a silicon crystal.
Dashed lines represent the dechanneling length $L_{\mathrm d}(0)$
for $\E=0.5$ and 20 GeV positrons in
single oriented silicon (110) crystal (see Table \ref{Si110.Table01})
}
\label{Latt_C-Si-Ge.fig}
 \end{figure}

\begin{itemize}
\item
In a CU the trajectory of a projectile particle follows the bending
profile.
This is possible when the electrostatic crystalline field
$U^{\prime}_{\max}$ exceeds the maximum centrifugal force
$F_{\rm cf}=\E/R$ due to the bending \cite{Tsyganov1976}.
For the bending profile (\ref{Geometry:eq.07b}) the maximum
curvature is $(1/R)_{\max}=4\pi^2a/\lambda_{\rm u}^2$.
Therefore, the  condition for \textit{stable channeling},
$C=F_{\rm cf}/U^{\prime}_{\max} \ll 1$,
takes the form shown by the first line
in (\ref{Restrictions_03:eq.01}).

\item 
It is implied that the CU operates in the \textit{Large-amplitude
regime}: $a> d$ ($d$ is the interplanar distance).
This limit, accompanied by the condition $C\ll 1$, is mostly
advantageous, since in this case the characteristic energies
of CUR and ChR are well separated.
A strong inequality $a\ll\lambda_{\mathrm{u}}$
ensures elastic deformation of the crystal.

\item 
\textit{Large number of periods} ensures that the radiation
emitted bears the features of the undulator radiation (narrow,
well-separated peaks in
the spectral-angular distribution of radiation).

\item
The last inequality in Eq. (\ref{Restrictions_03:eq.01}) suggests
that for effective operation of a CU its length $L_{\rm u}$ along
the incident beam direction must be chosen to be less than both
the dechanneling length
$L_{\mathrm d}(C)$ of the projectile particles and the photon
attenuation length $L_{\rm att}(\om)$.
These quantities stand for the spatial intervals over which the flux of
channeled particles (the quantity $L_{\mathrm d}(C)$) or the flux of
emitted photons (the quantity $L_{\rm att}(\om)$) decreases by a factor
of $e=2.718\dots$.

For positrons, a good estimation for $L_{\mathrm d}(C)$ can be obtained by
means of the following formulae \cite{BiryukovChesnokovKotovBook,ChannelingBook2014}:
\begin{eqnarray}
\begin{array}{l}
L_{\rm d}(C) = (1-C)^2 L_{\rm d}(0)\,,
\\
\displaystyle
L_{\rm d}(0)
=
{\E \over m_ec^2}\,
{256 \over 9\pi^2}\,
{a_{\rm TF} \over r_0}\,
{d \over \Lambda}\,.
\end{array}
\label{Restrictions_03:eq.02}
\end{eqnarray}
Here $L_{\rm d}(0)$ is the dechanneling length in the straight channel
(i.e. $C=0$), $r_0$ is the classical electron radius,
$Z$ and $a_{\rm TF}$ are the atomic
number and the Thomas-Fermi radius of the crystal atoms, respectively,
and
$\Lambda = 13.55 + 0.5 \ln\left(\E\, \mbox{[GeV]}\right) - 0.9 \ln(Z)$.

For the sake of reference, Table \ref{Si110.Table01} presents
the values of  $L_{\rm d}(0)$ calculated for
$\E=0.5$ and 20 GeV positrons in (110) planar channels
in silicon.

The attenuation length is related to the mass attenuation coefficient
$\mu(\omega)$: $L_a(\om)=1/\mu(\omega)$
(see, e.g., \cite{ParticleDataGroup2018}).
To calculate the values of $L_a(\om)$ in a wide range of photon
energies one can use the data on $\mu(\omega)$ available in
Refs. \cite{Henke,Hubbel}.

Figure \ref{Latt_C-Si-Ge.fig} shows the attenuation length versus
photon energy for a silicon single crystal.
For the sake of comparison, the dechanneling lengths $L_{\mathrm d}(0)$
for 0.5 and 20 GeV positrons
are also shown (dashed horizontal lines).

\end{itemize}

\begin{table}[h]
\caption{
Speed of sound $V_{l}$ (longitudinal, along the $\langle 100 \rangle$ direction),
interplanar distance $d$,
depth $U_0$ of the interplanar potential,
Lindhard's critical angle $\theta_{\rm L}(0)$
and dechanneling length
$L_{\mathrm d}(0)$
for $\E=0.5$ and 20 GeV  positrons in (110)
planar channels a silicon single crystal.
}
\footnotesize\rm
\centering
\begin{tabular}{p{2cm}p{2.cm}p{1.5cm}p{2cm}p{2.1cm}p{2.1cm}p{2.cm}p{2.cm}}
\hline
 &   & &  &\multicolumn{2}{c}{$\E=0.5$ GeV} &\multicolumn{2}{c}{$\E=20$ GeV} \\
$V_{l}$
& $d$
& $U_0$
& $U^{\prime}_{\max}$
&$\theta_{\rm L}(0)$  &$L_{\mathrm d}(0)$
&$\theta_{\rm L}(0)$   &$L_{\mathrm d}(0)$ \\
($10^5$ cm/s)
& (\AA)
& (eV)
& (GeV/cm)
& ($\mu$rad) & (cm)
& ($\mu$rad) & (cm) \\
8.43 \cite{Ioffe_DataBase} & 1.92 & 22  & 5.7
& 303    & 0.035 &  48  &   1.18  \\
\hline
\end{tabular}
\label{Si110.Table01}
\end{table}

\subsubsection{Restrictions specific for A-CU \label{Restrictions}}

\paragraph{Restrictions due to the finite time-of-flight. \label{Restrictions_02}}

The time-of-flight $\tau$ through the crystal of an ultra-relativistic projectile that
enters the crystal along the $z$ axis (see Fig. \ref{Figure:A-CU-scheme}) is calculated as
$\tau \approx \sqrt{2} L/c$.
The projectile will "see" a static bending pattern
(i.e. when the term $2\pi\nu_{\rm AW}t$ in the phase $\phi_0$
in Eq. (\ref{Geometry:eq.07b}) stays constant)
provided $2\pi\tau \ll 1/\nu  = {\lambda_{\rm AW} / V_{l}}$
where $V_{l}$ stands for the AW speed.
Written in terms of the undulator periods
$N={\sqrt{2} L / \lamu} = L/\lambda_{\rm AW}$
this inequality  reads:
\begin{eqnarray}
N \ll {3 \over 2\pi\sqrt{2}} {10^{10} \over V_{l}\, [\mbox{cm/s}]}\,.
\label{Restrictions_01:eq.04}
\end{eqnarray}
Using the value of $V_{l}$ presented in Table \ref{Si110.Table01}
one finds the following \textit{restrictions on the number of periods} in a silicon(110) based
A-CU:
\begin{eqnarray}
N
\ll
4\times 10^{3}\,.
\label{Restrictions_01:eq.06}
\end{eqnarray}

\paragraph{Restrictions due to the initial incident angle.
\label{Restrictions_03}}

Consider an ideally collimated beam entering an acoustically excited
crystal along the $z$ direction, see Eq. (\ref{Geometry:eq.07b}).
Different particles of the beam will "see"  different bending
profiles depending on the entrance point $Y_0$.
The incidence angle of
$\theta_i \approx |S^{\prime}(0)|=|ak_{\rm u}\sin(\phi_0)|$ with
respect to the tangent to the bent (110) plane
varies from $0$ up to $ak_{\rm u} = 2\pi a/\lamu$.
A particle can be accepted into the channeling mode if $\theta_i$ is
less than Lindhard's critical angle
$\theta_{\rm L}= \sqrt{2U_0 / \E}$ \cite{Lindhard},
where $U_0$ is the depth of the interplanar potential.
The values of $U_0$ and $\theta_{\rm L}(0)$ for $\E=0.5$ and 20 GeV
positrons in (110) planar channels in silicon crystal
are given in Table \ref{Si110.Table01}.

To ensure that all particles of the beam are accepted at the crystal
entrance the following condition must be met:
\begin{eqnarray}
\Bigl(\theta_i\Bigr)_{\max} = {2\pi a \over \lambda_{\rm u}} \leq \theta_{\rm L}\,.
\label{Restrictions_02:eq.01}
\end{eqnarray}

\paragraph{Restrictions due to the AW attenuation.}

The AW amplitude decreases with distance $x$ following the
exponential law $A_{\rm AW}(x) = A_{\rm AW}(0) \exp\Bigl(-\alpha x\Bigr)$,
where the attenuation coefficient $\alpha$ is measured in
 Nepers per unit length (e.g., Np/cm), where Neper stands for the
 natural logarithm of the ratio.
Any quantity, proportional to the AW amplitude (e.g., pressure)
follows the same exponential law.

A simple formula for the attenuation coefficient can be derived:
$\alpha = \kappa\,\nu^2$,
see, for example,
Refs. \cite{Landau7,LambRichter_ProcRoySocA_v293_p479_1966}.
Here the coefficient $\kappa= 2\pi^2  \eta / 2\rho V_l^3$ depends on the
viscosity tensor $\eta$, mass density $\rho$
and the AW velocity $V_l$.

\begin{figure}[h]
\centering
\includegraphics[scale=0.4,clip]{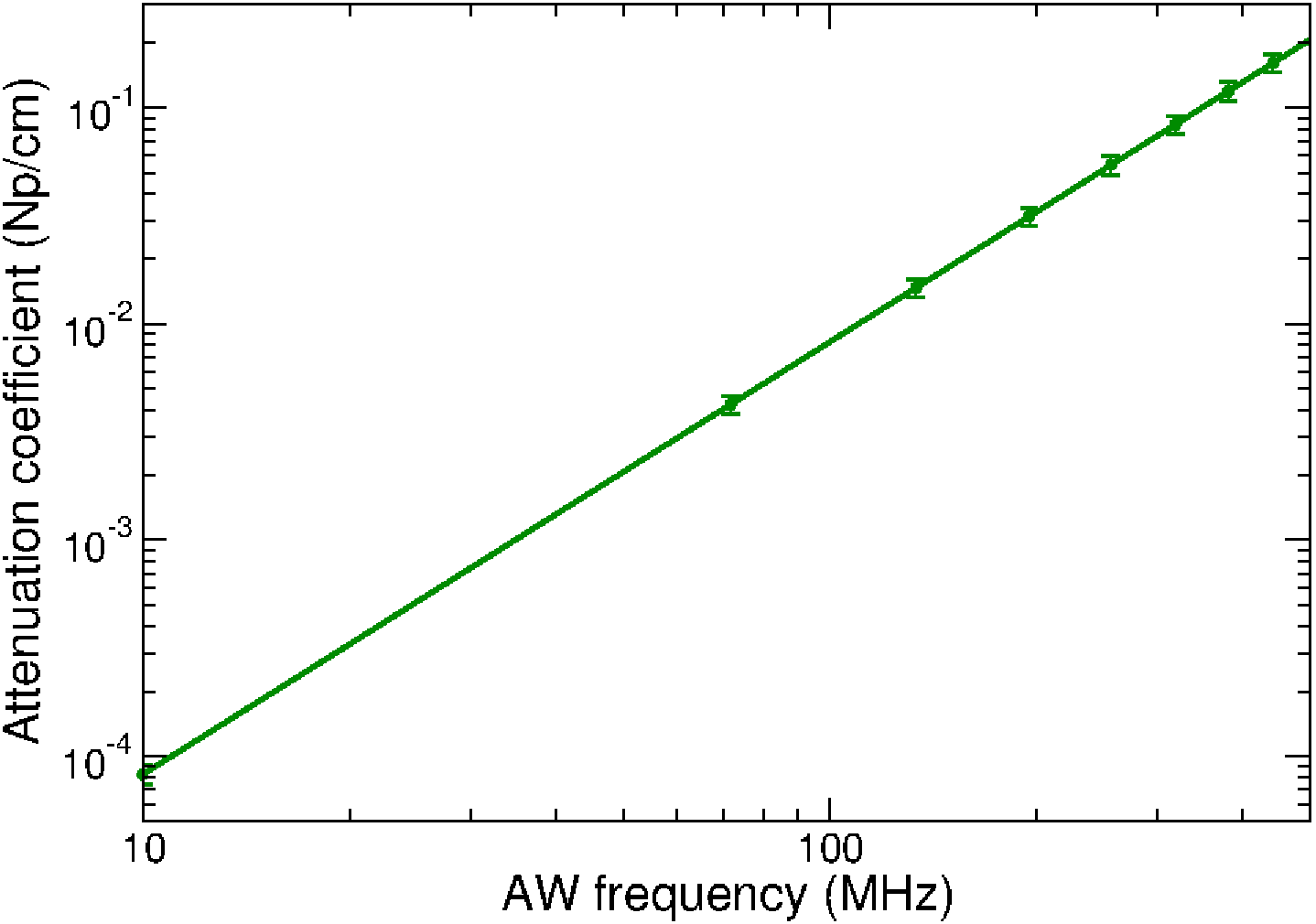}
\caption{
Attenuation coefficient  $\alpha$ (in Np/cm) versus AW frequency
for a longitudinal AW propagating along the $\langle 100 \rangle$
direction in a silicon single crystal.
}
\label{Attenuation_Si.fig}
 \end{figure}

In literature (see Ref.  \cite{HelmeKing_PhysStatSolA_v3_K33_1978}) one finds
$\alpha=0.87\pm 0.08$ Np/cm at $\nu=1.03$ GHz.
Using this value and taking into account that $\alpha$ scales as $\nu^2$ one can
calculate the attenuation coefficient in silicon for any AW frequency.
The resulting dependencies $\alpha(\nu)$ in the frequency range
$10 \dots 500$ MHz is shown in Fig. \ref{Attenuation_Si.fig}.

The restriction due to the AW attenuation can be formulated as follows
\begin{eqnarray}
\alpha L_v  \ll 1
\label{Attenuation:eq.06}
\end{eqnarray}
where $L_v$ is the crystal thickness along the $\langle 100 \rangle$
direction, see Fig. \ref{Figure:A-CU-scheme}.

\subsubsection{Case studies \label{Ranges}}

For a given energy of a projectile positron and a given crystal,
conditions (\ref{Restrictions_03:eq.01}),
(\ref{Restrictions_01:eq.06}), (\ref{Restrictions_02:eq.01}) and
(\ref{Attenuation:eq.06}) define the ranges of parameters,
which include crystal thickness, AW frequency, bending amplitude and
period, within which the acoustically excited CU can be considered
as a feasible scheme.
Such an analysis has been carried out for $\E=0.5$, $5$, $10$ and
$20$ GeV positron channelling in (110) planar channels in single
diamond, silicon and germanium crystals
\cite{KorolSolovyov2023_unpublished}.

\begin{figure}[h]
\centering
\includegraphics[scale=0.3,clip]{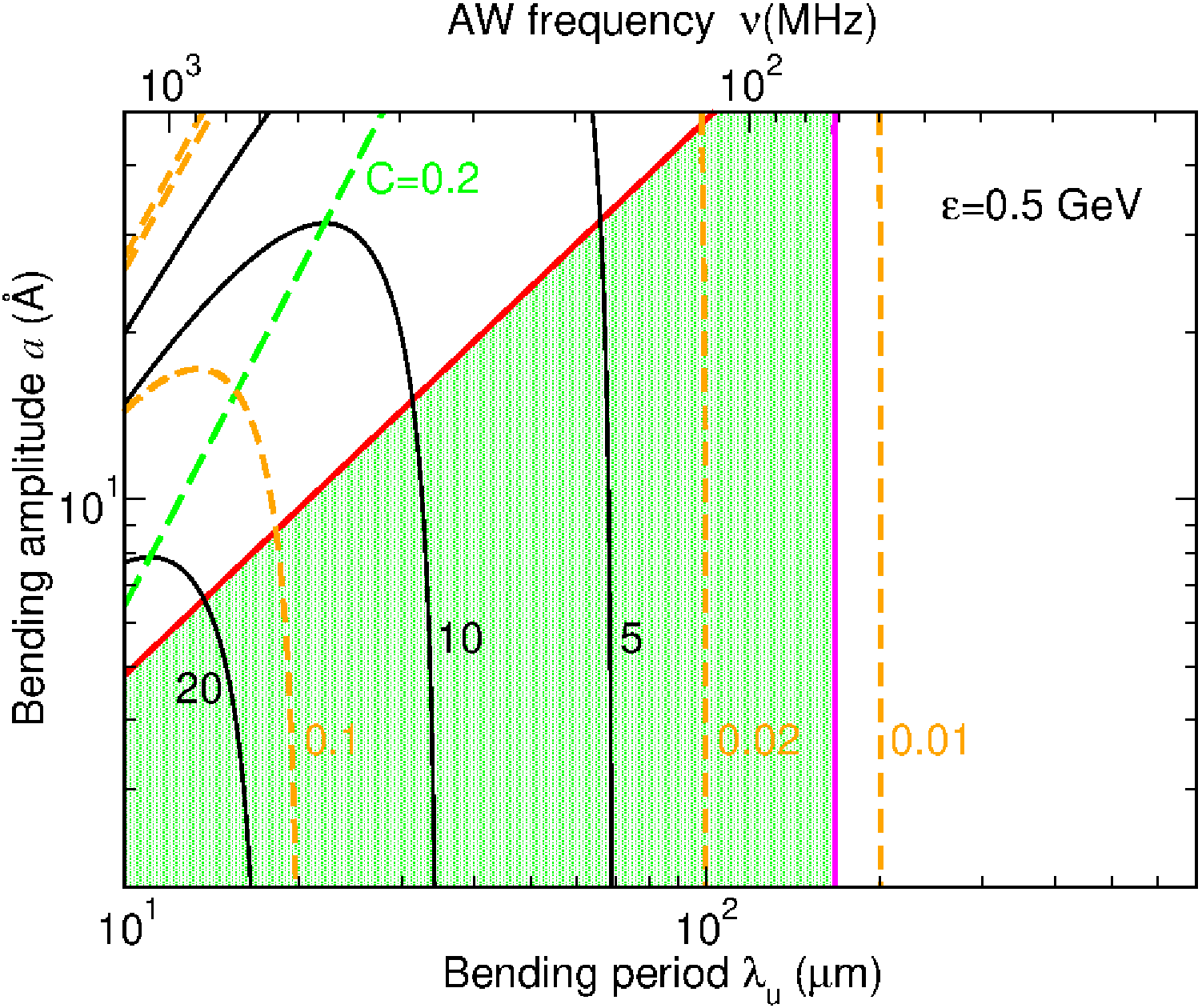}\hspace*{0.2cm}
\includegraphics[scale=0.3,clip]{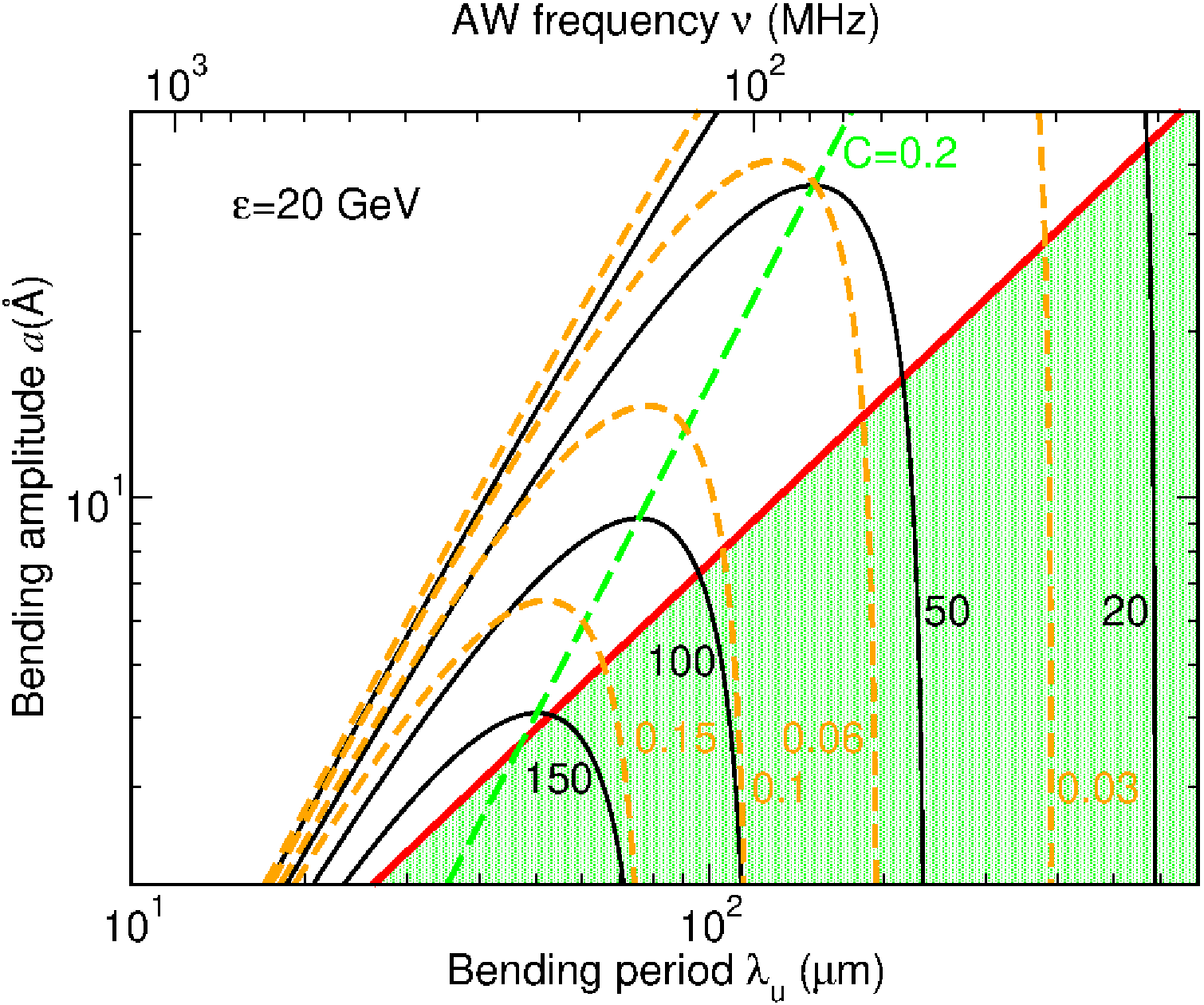}
\caption{
Ranges of bending period $\lamu$ and amplitude $a>d=1.92$ \AA{}
to be probed to construct a Si(110)-based A-CU.
The data refer to $\E=0.5$ GeV (left graph) and $20$ GeV (right
graph) positrons.
For each energy, the CU can be considered in the (shadowed) domain
lying below the red line, which corresponds to the
condition (\ref{Restrictions_02:eq.01}).
In the case of $\E=0.5$ GeV additional restriction is due to the
photon attenuation: the magenta line marks corresponds to
$L_{\mathrm d}(C)=L_{\rm att}(\om)$, so that the favourable region,
$L_{\mathrm d}(C)>L_{\rm att}(\om)$, lies to the left of the line.
The green dashed line indicates the value $C=0.2$;
Integers indicate the number of undulator periods within the
dechanneling length: $\Nd = L_{\mathrm d}(C)/\lamu$,
see Eq.  (\ref{Restrictions_03:eq.01}).
Dashed (orange) lines denotes the contours $\alpha L_{\mathrm d}(C)=const$
for the indicated value of $const$.
}
\label{AW_ranges_Si110.fig}
 \end{figure}

Here we present two case studies that refer to $\E=0.5$ GeV
(MAMI energies, Ref. \cite{BackeEtAl_EPJD_v76_150_2022}) and $20$ GeV
(CERN energy, Ref. \cite{CERN})
positrons channeling in acoustically excited silicon crystals.
Figure \ref{AW_ranges_Si110.fig}  shows the domains of $\lamu$ and $a\geq d$ \AA{}
 (the shadowed areas) within which the conditions
(\ref{Restrictions_01:eq.06}), (\ref{Restrictions_03:eq.01}) and (\ref{Restrictions_02:eq.01}) are fulfilled for
$\E=0.5$ GeV (left graph)  and $20$ GeV (right graph) positrons.
In each graph the red line denotes the boundary
$a = \theta_{\rm L}(C)\lamu /2\pi$ so that the condition
(\ref{Restrictions_02:eq.01})
is fulfilled below the line.
The solid magenta line, shown in the graph for
$\E=0.5$ GeV, corresponds to the boundary $L_{\mathrm d}(C)=L_{\rm att}(\om)$.

\subsection{FEM simulations of lattice structural modulation
\label{Methods_FEM}}

In this section we present the computational methodology
used to simulate the acoustically induced deformation of the Si
crystal lattice using the Finite Element Method (FEM) by means
of the  LSDYNA software \cite{LSDYNA}.

\begin{figure}[h]
    \centering
    \includegraphics[width=1\linewidth]{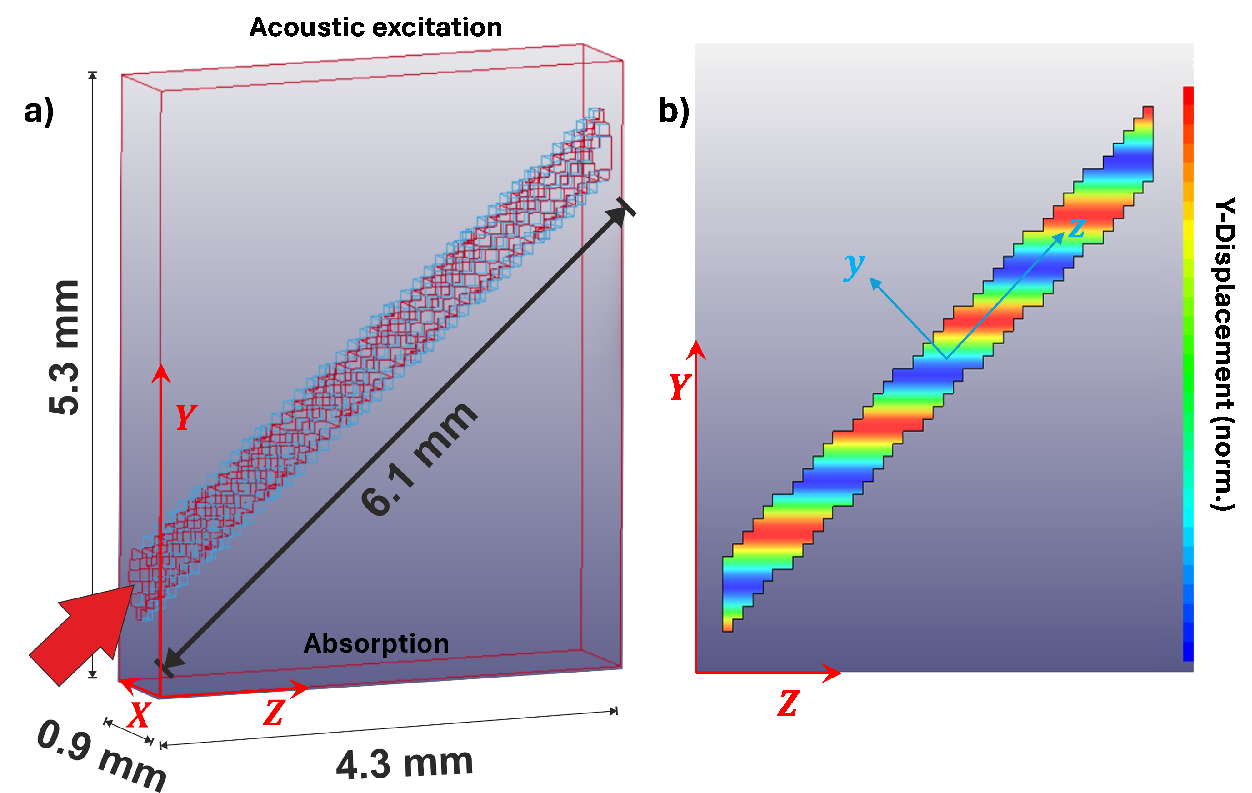}%
    \caption{\textit{Panel a):} Silicon crystal domain used for the
    FEM simulations of the periodic lattice deformation due to
    the acoustic field.
    The red arrow shows the direction of the incident positron beam.
    \textit{Panel b):}
    Derivation of the channel displacement perpendicular to the
    incident beam direction.
    Colour bar: red and blue indicate the maximum positive and
    negative pressure, respectively.}
    \label{Figure:Crystal-domain}
\end{figure}

An A-CU based on the AOM layout suitable for the channeling of
$\E= 20$ GeV positrons is considered as a case study.
Such positron beams are currently available at the facilities of the 
European Organization for Nuclear Research (CERN)
\cite{CERN}.
A silicon single crystal considered has the
following dimensions (see Fig. \ref{Figure:Crystal-domain}a) for the
reference frames adopted):
$Z_{\rm c}=4.3$ mm, $Y_{\rm c}=5.3$ mm and $X_{\rm c}=0.9$ mm.
The ultra-relativistic positron beam is assumed to traverse the crystal
along the (110) planar direction $z$.
Therefore, the thickness along the incident beam is
$L=\sqrt{2} Z_{\rm c} = 6.1$ mm,
which is approximately two times less than the dechanneling length
$L_{\rm d}(0)$ calculated from
Eq. (\ref{Restrictions_03:eq.02})  (see also Table \ref{Si110.Table01}).

The domain is discretised with a spatial resolution of
50 \textmu m $\times$ 50 \textmu m $\times$ 9 \textmu m.
A mesh convergence analysis is performed and the total number of
elements used for the simulations is approximately one million.
The applied harmonic pressure on the upper $XZ$ plane has an
amplitude of 4 MPa.
The 10 MHz and 40 MHz frequency cases are considered with free
boundaries on the two $YZ$ and two $XY$ planes and with
non-reflecting (absorber)
boundary conditions on the lower (100) $XZ$ plane.
The free boundaries on the peripheral planes allow for the crystal
to deform in all directions, while the non-reflective boundary at the
bottom $XZ$ plane
eliminates reflections and thus the accumulation of acoustic energy
and the formation of standing waves in the crystal.

Analysis of the FEM results is done by considering diagonal paths
along the $z$ direction inside the crystal domain
(see Fig. \ref{Figure:A-CU-scheme}).
More specifically, the positron beam is considered to penetrate the
crystal at $\theta=45^{\circ}$ at the left $XY$ plane in the middle
of the $X$ dimension.
For each node, the $\rho_Z$ and $\rho_Y$ displacements in the $Z$
and $Y$ directions, respectively, are extracted and transformed
into the displacement $\rho_{\perp}$ perpendicular to the direction
of the incident beam by means of the following transformation:
\begin{equation}
\rho_{\perp} = \rho_Z \sin\theta + \rho_Y \cos\theta\,.
\label{eq:FEM_profile}
\end{equation}
This relation allows one to calculate the bending profile of the
deformed channels along the diagonal direction which starts at the
crystal entrance, i.e. at the points $(X,Y)$ at $Z=0$,
see Fig. \ref{Figure:Crystal-domain}b).

\subsection{Calculation of trajectories and light emission
\label{MBN_Methodology}}

Numerical modelling of the passage of ultra-relativistic positrons
in a crystalline environment and of the related phenomena
have been carried out by means of the commercially available
multi-purpose scientific computer package \MBNExplorer
\cite{MBNExplorer_2012,MBNExplorer_site}
and a supplementary  special multitask software toolkit \MBNStudio
\cite{SushkoEtAl_2019-MBNStudio,MBNExplorer_site}.
The simulation procedure is based on the formalism of classical
relativistic molecular mechanics (Rel-MD) and describes the motion
of a projectile particle in the laboratory reference frame along
with dynamical simulations of the environment
\cite{MBN_ChannelingPaper_2013}.
\MBNExplorer solves the following relativistic equations of motion
(see, e.g., \S 17 in Ref. \cite{Landau2}):
\begin{eqnarray}
\left\{\begin{array}{l}
\displaystyle{
\dot{\bfv} =
{1 \over m \gamma}
\left(\bfF -  {\bfv\left(\bfF \cdot \bfv\right) \over c^2}\right)
}
\\
\dot{\bfr} = \bfv
\end{array}
\right. \ ,
\label{Equations:eq.01}
\end{eqnarray}
where $\gamma = \E/mc^2 =\left(1- v^2/c^2\right)^{-1/2}$
is the relativistic Lorentz factor of a projectile of energy $\E$
and mass $m$,
with $c$ being the speed of light.

In the most general case, the force $\textbf{F}$ is calculated as
the sum of two terms:
$\mathbf{F} = \mathbf{F}_{\rm em} + \mathbf{F}_{\rm rr}$.
Here $\mathbf{F}_{\rm em}$ stands for the total electromagnetic
force due to (i) the electrostatic field $\mathbf{E}$ created by
atoms of the medium and/or by external sources of electric field,
and (ii) external magnetic field $\mathbf{B}$.
The term $\mathbf{F}_{\rm rr}$ stands for the radiative reaction force
\cite{Landau2,SushkoEtAl:NIMB_v535_117_2023}.
In the current simulations, only the electrostatic interaction of
the projectile positron with crystal atoms has been accounted for.
Hence, the force $\mathbf{F}$ was calculated as
$\bfF = q\mathbf{E}(\mathbf{r})$ where $q$ is the positron charge.
The electric field is calculated via
$\mathbf{E}(\mathbf{r}) = - {\partial \phi(r) / \partial \mathbf{r}}$
where  $\phi({\textbf{r}})$ denotes the electrostatic potential
created by the crystal atoms in the point $\textbf{r}$:
\begin{eqnarray}
\phi(\textbf{r})
=
\sum_{j} \phi_{\mathrm{at}}\left(\left|\textbf{r} - \textbf{r}_j\right|\right)\,.
\label{sec:relMD_01:eq.07}
\end{eqnarray}
Here $\phi_{\mathrm{at}}$ are the potentials of individual atoms and
$\textbf{r}_j$ denotes the position vector of the $j$th atom.
In the simulations the atomic potentials were considered within the
Moli\`{e}re approximation \cite{Moliere}.
 Due to a rapid decrease of $\phi_{\mathrm{at}}$ with increasing
 the distance from an atom, the sum can be truncated in practical
 calculations.
 Only atoms located inside a sphere of the (specified) cut-off
radius $\rho_{\max}$ with the center at the current location of the
projectile particle.

Statistical independence of the simulated trajectories is due to
 several features that are implemented in the code.
Firstly, the transverse coordinates and velocities of a projectile
at the crystal entrance are generated randomly accounting for their
distribution determined by the transverse size and divergence of the
beam (which are the input parameters).
Therefore, for each trajectory the integration of the equations of
motion (\ref{Equations:eq.01}) starts with randomly different
initial conditions.
Secondly, in the course of a trajectory simulation the positions
of the crystal atoms are generated on the fly accounting for random
displacement from the nodes due to thermal vibrations.
Finally, \textsc{MBN Explorer} allows one to to account for random
events of inelastic scattering of a projectile particle from
individual atoms while integrating the classical equations of motion
\cite{SushkoEtAl:arXiv_2405.07633}.
These events result in random change in the direction of the
projectile motion.
More details on the algorithms implemented in \textsc{MBN Explorer}
to compute trajectories of the particles passing through a medium
are presented in
\cite{MBN_ChannelingPaper_2013,KorolSushkoSolovyov:EPJD_v75_p107_2021,%
SushkoEtAl:NIMB_v535_117_2023,SushkoEtAl:arXiv_2405.07633}.

As a direct consequence of the aforementioned features, each simulated
trajectory corresponds to a unique crystalline environment and,
therefore, all simulated trajectories are statistically independent
and can be analyzed further to quantify the channeling
process as well as the emitted radiation.
The averaged spectral distribution of energy radiated within the cone
$\theta \leq \theta_0 \ll 1$ with respect to the incident velocity is
calculated as follows:
\begin{eqnarray}
\left\langle
\frac{\d E(\theta _0)}{\d\omega}\right\rangle
=
\frac{1}{N} \sum ^N_{n=1}\int _0^{2\pi}d\phi \int _0^{\theta _0}\theta \d\theta
\frac{\d^3 E_n}{\d\omega \d\Omega }\ .
\label{eq:02} 
\end{eqnarray}
Here, $\omega $ stands for the frequency of radiation;
$\Omega =(\theta, \phi)$ is the solid angle of the photon emission;
$\d^3 E_n/\d\omega \d\Omega$ is the spectral-angular
distribution of the radiation emitted by a projectile
moving along the $n$th trajectory; and $N$ stands for the total
number of the simulated trajectories.
The numerical computation of $\d^3 E_n/\d\omega \d\Omega$ is
based on the quasi-classical method \cite{Baier} which
combines the classical description of the motion of a particle in an
external field with the quantum corrections due to the radiative
recoil.
The sum is carried out over all simulated trajectories.
Therefore,
The averaged spectrum therefore takes into account account
all types of radiation emitted in the crystalline target of
a given geometry (these
include channeling radiation, coherent and incoherent bremsstrahlung,
crystalline undulator radiation).

\section{Results  \label{Results}}

\subsection{Crystal lattice structural modulation}

\begin{figure}[h]
    \centering
    \includegraphics[width=0.97\linewidth]{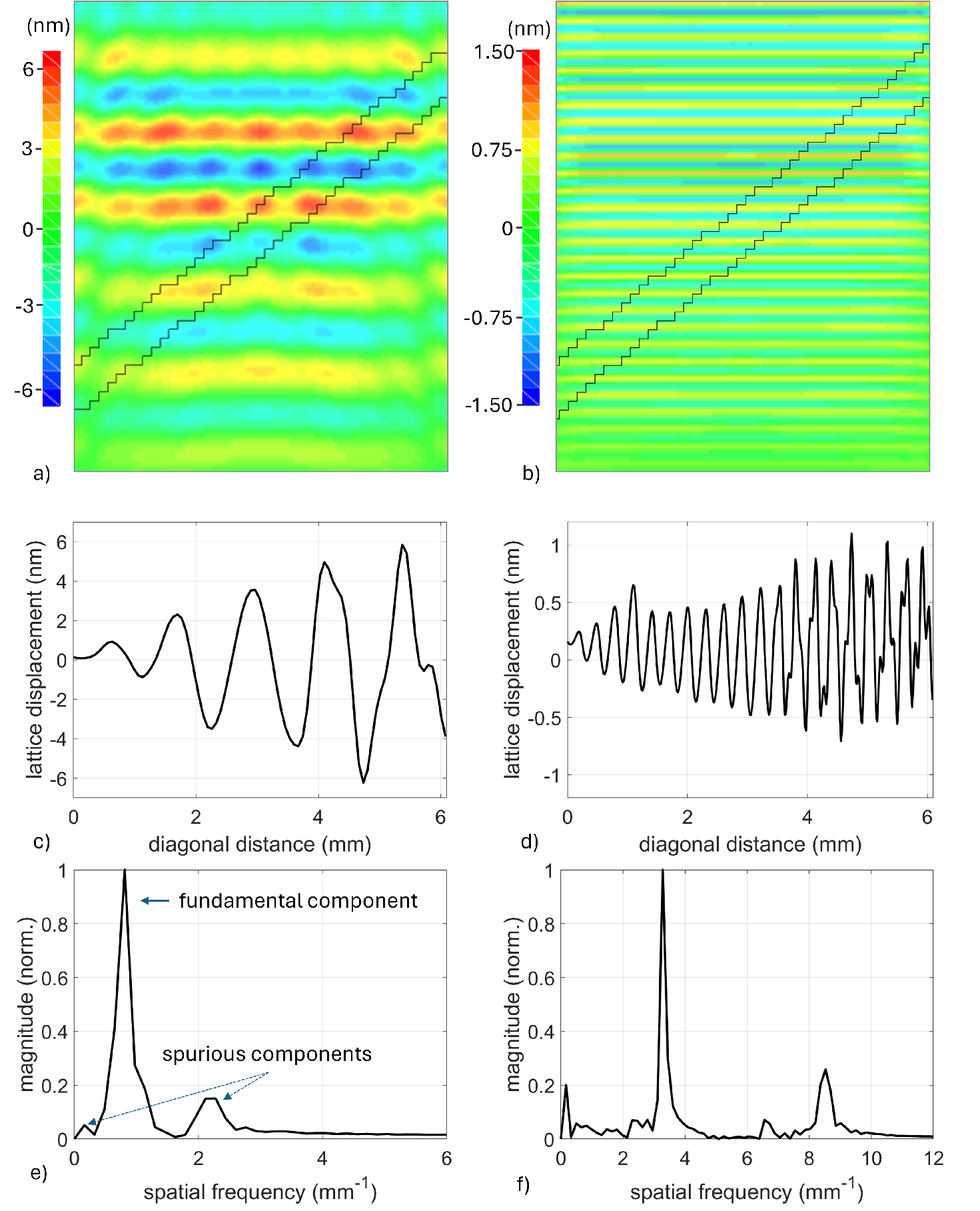}
    \caption{
    \textit{Graphs a) and b):}
  Colour maps of the crystal lattice displacement in the $Y$ direction.
 \textit{Graphs c) and d):} The corresponding bending profiles $y=y(z)$.
 \textit{Graphs e) and f):} The lattice modulation spectra obtained
 by means of the FFT of the bending profiles.
 The left and right columns refer to $\nu =10 $ and 40 MHz,
 respectively.
    }
    \label{Figure:Bending-profiles}
\end{figure}

Figures \ref{Figure:Bending-profiles}a) and b) show the colour maps
of the simulated Y-displacement within the Si crystal domain, generated
by the 4 MPa sinusoidal excitation at frequencies
$\nu=10$ MHz and 40 MHz, respectively.
The bending profiles along the diagonal direction, indicated by the
black lines, are chosen to be far from the upper and lower boundaries
to ensure maximum homogeneity of the acoustic field.
Figures \ref{Figure:Bending-profiles}c) and d) show the
bending profiles extracted from the displacement maps using the
transformation (\ref{eq:FEM_profile}).
Figures \ref{Figure:Bending-profiles}e) and f) show the spatial
frequency spectrum obtained from the Fast Fourier Transform (FFT)
of the profiles.
There are 5 periods along the diagonal path for the 10 MHz
excitation with $\lambda_{\rm u}=1.2$ mm wavelength and 20 periods
for the 40 MHz excitation with $\lambda_{\rm u}=0.3$ mm.
The lattice shift is much larger for the 10 MHz excitation,
peaking at 6 nm, while at 40 MHz the maximum shift
is limited to 1.2 nm.

From Fig. \ref{Figure:Bending-profiles} it can be seen that the bending
profiles are characterised by a strong fundamental harmonic component,
together with spurious components that introduce anharmonicity into the
profile.
In both cases of excitation, spurious spectral components appear
both below and above the fundamental frequency, referred to as
low and high frequency components.
The low frequency components are at the same frequency for both the
$\nu=10$ and 40 MHz and can be related to a bulk compression of the
material in the $Y$ direction, caused by the excitation pressure.
In addition, the high frequency spurious components can be attributed to
the "breathing" of the crystal in the $X$ and $Z$ directions caused by the
pressure excitation.
Compression in the $Y$ direction leads to a lateral expansion of the
crystal, inducing perturbations along the $X$ and $Z$ directions.
Such an effect can be clearly observed, e.g., in
Fig. \ref{Figure:Bending-profiles}a),
where nodes and anti-nodes appear in the displacement along the
$Z$ direction.

The focus of the present work is on the case of the free crystal,
since this scheme is is easier to implement experimentally.
Nevertheless, additional test simulations have been carried out for
a crystal constrained with respect to lateral deformation.
For such behaviour, simply supported plate boundary conditions were
applied to the two $YZ$ planes and the two $XY$ planes,
restricting lattice displacement in the the $X$ and $Z$ directions.
A harmonic pressure  of 4 MPa and 10 MHz was applied on the upper
$XZ$ plane and non-reflective (absorber) boundary conditions were
applied at the lower $XZ$ plane.
The simulations showed that restricting the lateral deformation of
the crystal leads to the complete elimination of the spurious high
frequency components, supporting the above attribution of their
origin to the "breathing" of the crystal.
Therefore, further investigation of the experimental realisation of the
simply supported case is of interest due to the expected higher
efficiency in the generation of near monochromatic gamma
radiation.

\subsection{Rel-MD simulations trajectories \label{MBN_Trajectories}}

\begin{figure}[h]
\centering
\includegraphics[scale=0.305,clip]{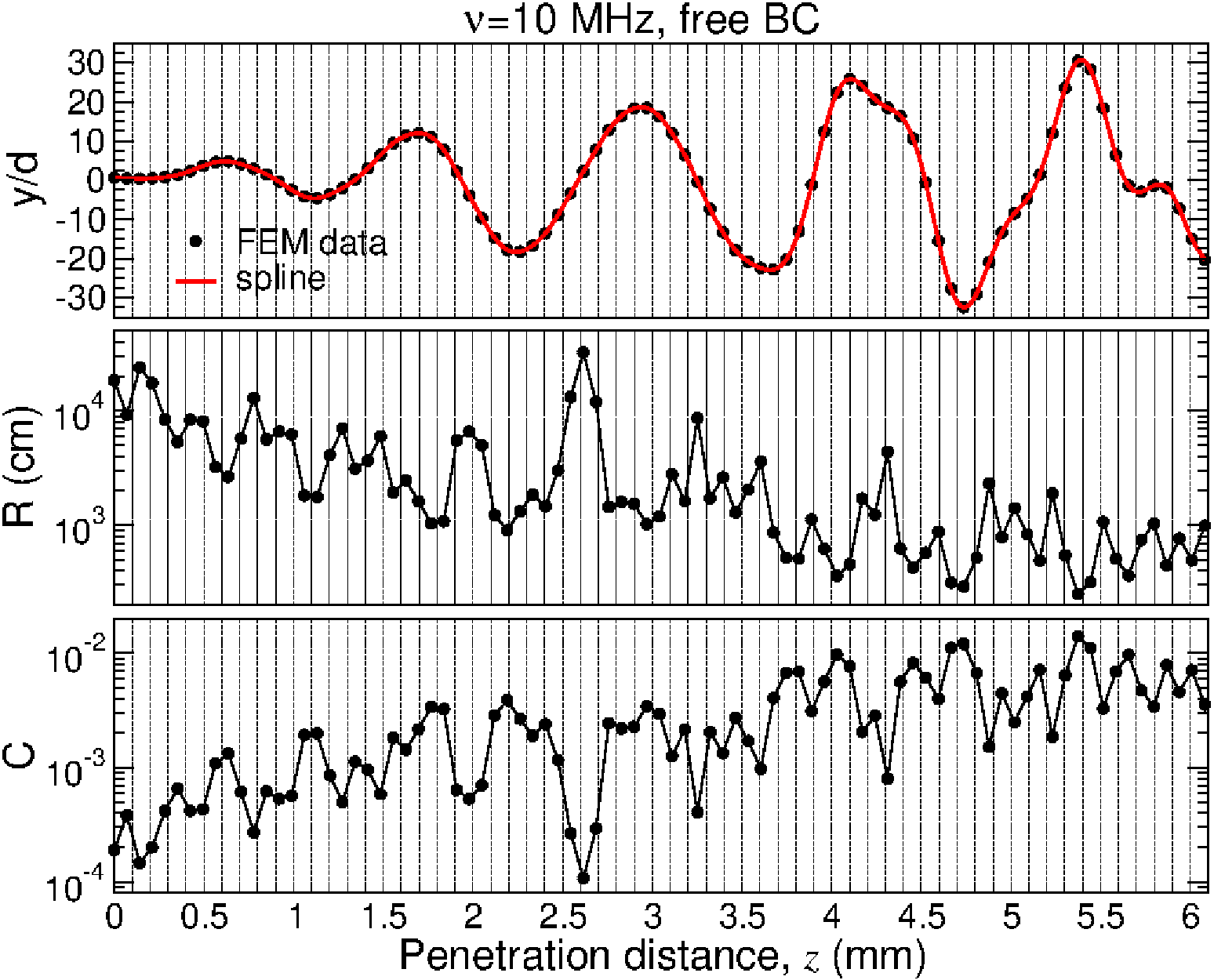}\hspace*{0.5cm}
\includegraphics[scale=0.305,clip]{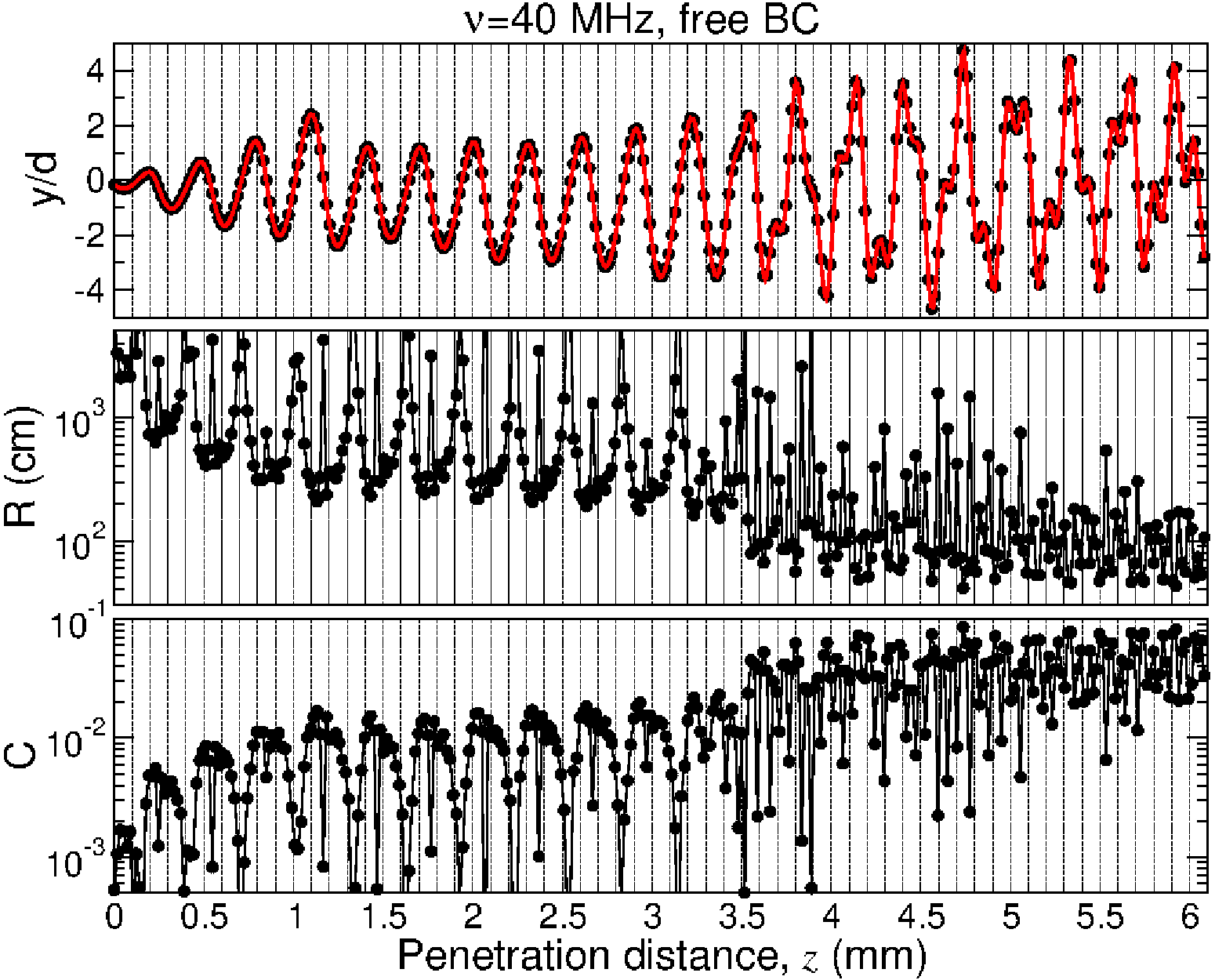}
\caption{
\textit{Top row.} Bending profiles of (110) planes in Si due to AW
with free boundary conditions: FEM results (symbols) connected with
a smooth spline.
\textit{Middle row.} Values of the local curvature radius
$R(z)\approx \left(|y^{\prime\prime}_{zz}|\right)^{-1}$.
\textit{Bottom row.}: Bending parameter C=centrifugal force / interplanar
force.
\textit{Left column} corresponds to the AW frequency 10 MHz, \textit{right
column} – to 40 MHz.
The corresponding values of bending periods are $\lambda_{\rm u}=1$
and 0.25 mm.
}
\label{Profiles_free_RC.fig}
 \end{figure}

The FEM modelling of the lattice structural modulation shows
deviation of the resulting periodic bending of the (110) planes from the
ideal harmonic profile defined by Eq. (\ref{Geometry:eq.07b}),
For the sake of reference, these profiles are reproduced in the top
graphs in Fig. \ref{Profiles_free_RC.fig} where the solid smooth
lines stand for the splines obtained from the FEM data (symbols).
Note that the transverse coordinate $y$ is scaled by the interplanar
distance $d$, see Table \ref{Si110.Table01}.
The middle row in the figure shows the variation of the curvature radius
$R(z)=(1+y^{\prime2}_{z})^{3/2}/|y^{\prime\prime}_{zz}|
\approx 1/|y^{\prime\prime}_{zz}|$ along the profile.
It is seen that this quantity greatly exceeds Tsyganov's
critical radius $R_{\min}=\E/U_{\max}^{\prime} \approx 3.5$ cm, which
results in small values of the bending parameter
$C(z)=\E/U_{\max}^{\prime}(R(z))=R_{\min}/R(z)$ in all points of the
profile (bottom row).

\begin{figure}[ht]
\centering
\includegraphics[scale=0.305,clip]{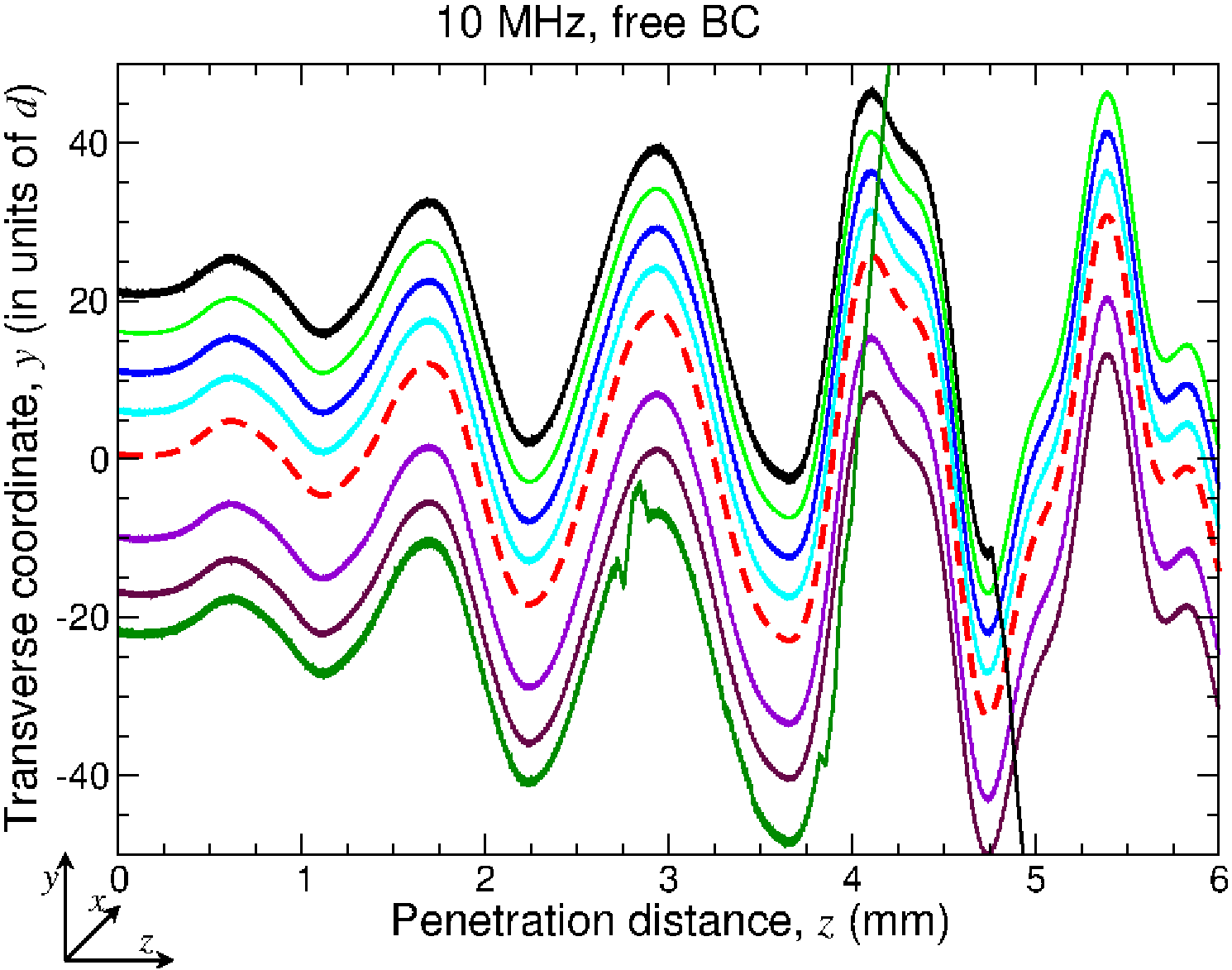}\hspace*{0.5cm}
\includegraphics[scale=0.305,clip]{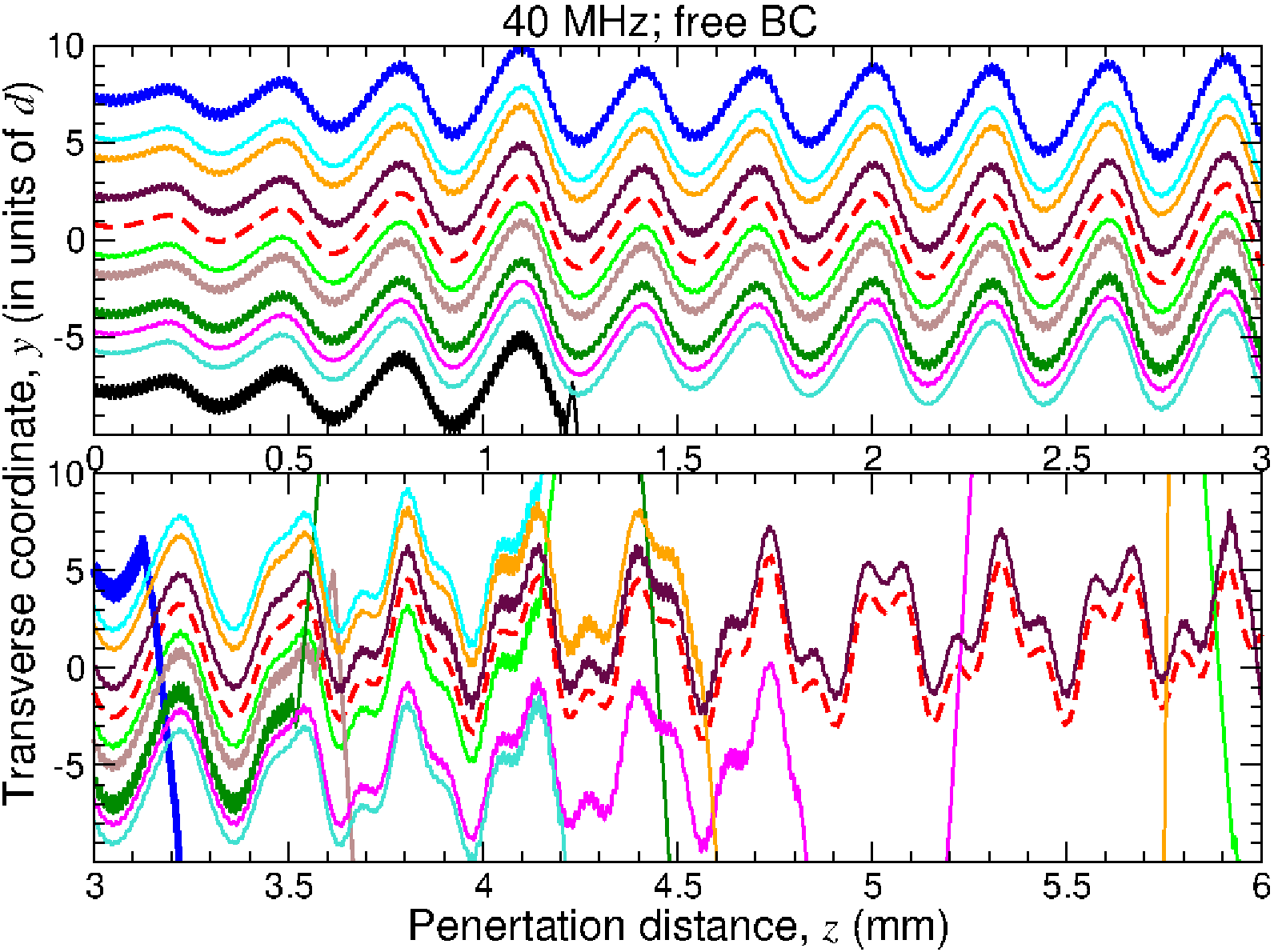}
\caption{
Exemplary trajectories (solid lines) of 20 GeV positrons incident on
6 mm thick acoustically excited Si crystal.
The $z$-axis of the reference frame is directed along the
incoming projectiles, the $(xz)$-plane is parallel to the (110)
crystallographic planes in the linear crystal, and the $y$-axis is
perpendicular to the planes.
The Si(110) interplanar distance is $d = 1.92$ \AA.
The solid lines represent the projections of the trajectories
on the $(yz)$-plane.
Left and right columns correspond to the bending profiles shown in Fig.
\ref{Profiles_free_RC.fig} (drawn in dashed lines in this figure).
For the sake of clarity and convenience of the visualization, the
trajectories in the right figure are shown in two segments:
from the entrance point $z= 0$ up to $z=3$ mm - the upper panel;
from $z=3$ up to 6 mm – the lower panel.
}
\label{Trajectories_free_RC.fig}
 \end{figure}


 A large number (about $10^4$) of trajectories of 20 GeV
 positrons were simulated for each of the profiles.
 Figure \ref{Trajectories_free_RC.fig} presents several trajectories
 of particles passing through a 6 mm thick silicon crystal being
 incident along the (110) planar direction (unbent) at the crystal
 entrance.
 The left and right graph correspond to the bending profiles (dashed
 lines) obtained by propagating acoustic waves of 10 and 40 MHz along
 the $\langle100\rangle$ axial direction.
In the case of the 10 MHz profile, a large number (ca 50 per cent)
of the incident particles pass through the whole crystal moving in the
channeling mode.
For a less regular and smooth 40 MHz profile the statistics is
different: approximately 60 per cent of the particle channel through the
first 3 mm of the crystal and only few percent continue channeling up to
the end of the crystal.
A channeling particle experiences (i) channeling oscillations while
moving along a
periodically bent centerline, and (ii) stochastic motion
along the $x$ axis due to the multiple scattering from
crystal constituents.
The latter motion is not visualized by Fig. \ref{Trajectories_free_RC.fig}
since it shows the 2D projections of the trajectories on the $(yz)$-plane.
The channeling oscillations, whose amplitude is subject to the condition
$a_{\rm ch} < d/2$, are more distinguishable against the 40 MHz profile,
since in this case the average bending amplitude $\langle a\rangle \approx 2.5d$ is much smaller
than for the 10 MHz profile where $\langle a\rangle \approx 19d$.
For both bending profiles the (average) period $\langle \lambda_{\rm ch}\rangle$
of the channeling oscillations is approximately equal to 12.7 $\mu$m.
This value correlates with the estimate which can be derived using the harmonic
approximation for the continuous interplanar potential \cite{Lindhard} in Si(110):
$\lambda_{\rm ch} = \pi d (\E/2U_0) \approx 13$ $\mu$m (see Table \ref{Si110.Table01}
for the $d$ and $U_0$ values).
For both profiles $\lambda_{\rm ch} \ll \lambda_{\rm u}$.

\subsection{Spectral-angular distribution of the emitted radiation
\label{MBN_Emission}}
%

\begin{figure}
\centering
\includegraphics[width=0.85\linewidth]{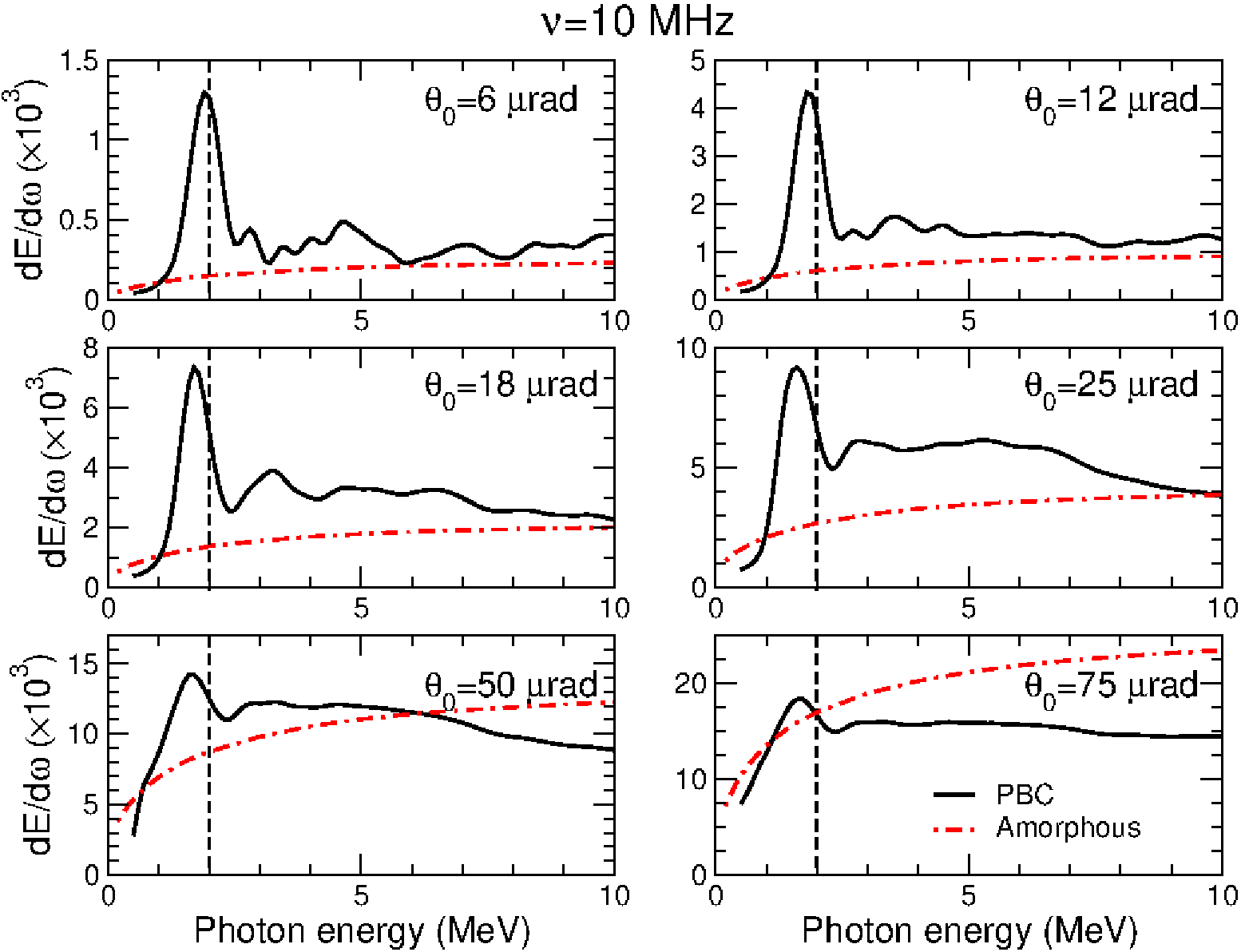}
\caption{
Spectral distribution $\d E(\theta_0)/\d(\hbar \om)$ of
radiation emitted by a 20 GeV positron beam in the 10 MHz A-CU
(black solid curves).
Six graphs correspond to different emission cones $\theta$ (as indicated)
along the incident beam direction.
Vertical dashed lines mark the position of the fundamental harmonic
(in the forward direction, $\theta=0$) estimated for the ideal undulator
(see explanations in the text).
For the sake of comparison, the radiation generated in the amorphous 6 mm
thick silicon target is also shown, red dash-dotted curves.
}
\label{10f_Spectra.fig}
\end{figure}

\begin{figure}
\centering
\includegraphics[width=0.85\linewidth]{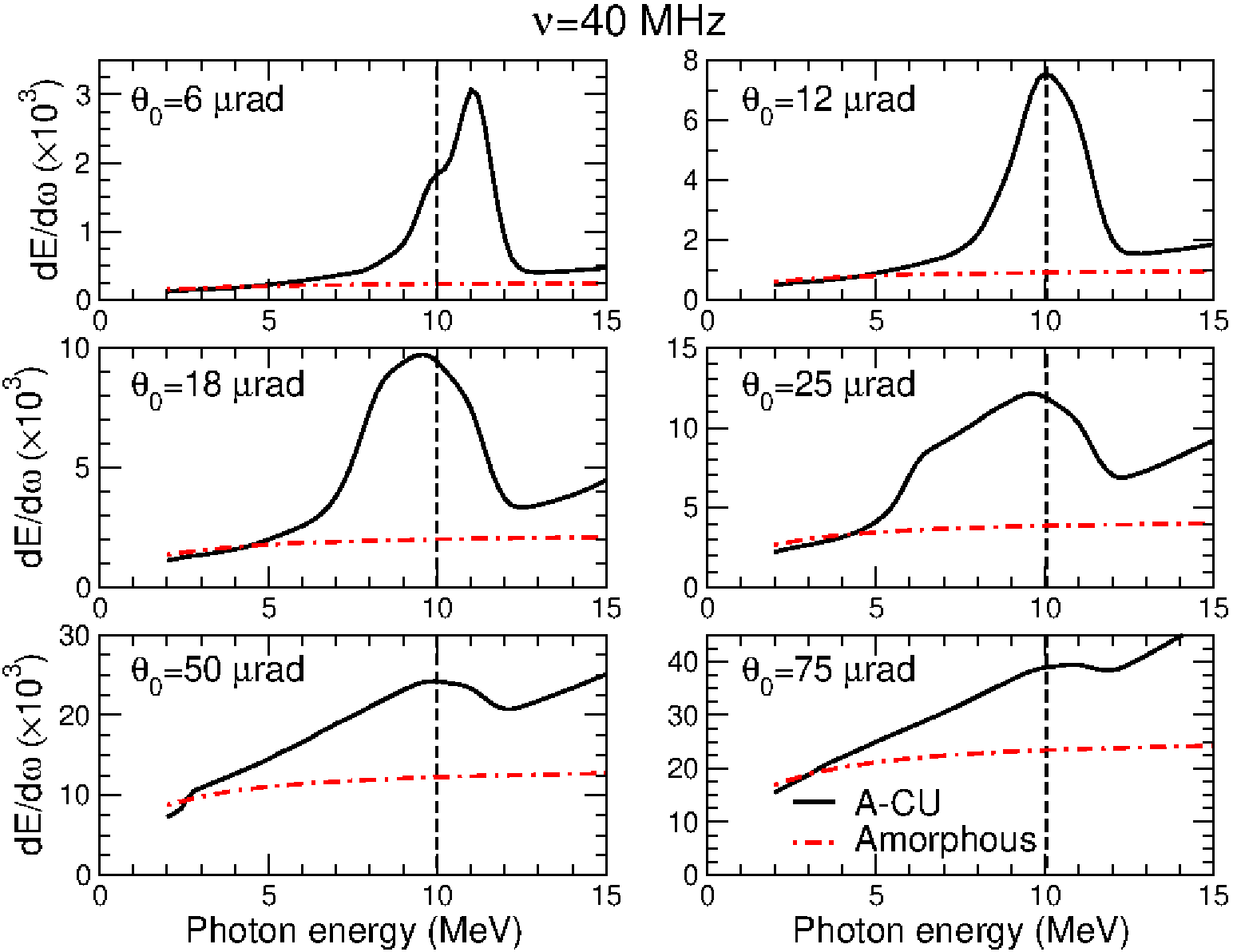}
 \caption{
Same as in Fig. \ref{10f_Spectra.fig} but
 for the $\nu = 40$ MHz A-CU.
}
 \label{40f_Spectra.fig}
\end{figure}

For each trajectory simulated ($n=1,\dots,N$) the spectral-angular distribution
of radiation $\d E^3_n/\d\om \d \Om$ has been calculated numerically.
The data obtained was further used in Eq. (\ref{eq:02}) to calculate the averaged
spectral distributions for several emission cones $\theta_0$ along the incident beam direction.
Solid black curves in Figs. \ref{10f_Spectra.fig} and
\ref{40f_Spectra.fig} present the results of these calculations
for the 10 and 40 MHz A-CU, respectively.
In each figure, six panels show the distributions computed
for the specified values of $\theta_0$.
Note that $\theta_0=25$ $\mu$rad corresponds to the natural
emission cone $\gamma^{-1}$ for a 20 GeV positron.
The curves drawn in dash-dotted red lines show the spectral
distributions of radiation emitted in the amorphous 6 mm thick silicon
target.
To simulate the amorphous environment the procedure that had been
described and utilized previously
\cite{MBN_ChannelingPaper_2013,KorolSushkoSolovyov:EPJD_v75_p107_2021}
was used.
In short, the procedure is as follows.
As mentioned in Sect. \ref{MBN_Methodology} above, in a
\textit{crystalline} medium the position vectors of atomic nuclei
are generated with account for random displacement from the nodes
due to thermal vibrations corresponding to a specified temperature $T$.
By introducing unrealistically large values of the root-mean-square
thermal vibration amplitude $u_T$ (for example, comparable to the
lattice constant) it is possible to generate sufficiently large
displacements, so that the resulting structure will
mimic the \textit{amorphous} medium.
In the current simulations the amplitude $u_T$ was set to 2 \AA.
Taking into account that a unit cell of a silicon crystal (with the
lattice constant equal to 5.43 \AA) contains 8 atoms, the quoted value
of $u_T$ ensures randomness of atomic positions in the sample.

Comparing the solid and dash-dotted curves in Figs.
\ref{10f_Spectra.fig}  and \ref{40f_Spectra.fig} one concludes
that for small emission cones,
$\theta_0 \leq \gamma^{-1} \approx 25$ $\mu$rad, the acoustically
excited crystals deliver a strong enhancement of the gamma-ray
radiation in a narrow spectral band
peaked at approximately $\hbar\om_{\max} \approx 1.5 - 2$ MeV for
the 10 MHz excitation and at $\hbar\om_{\max} \approx 9 - 11$
MeV for $\nu = 40$ MHz over the incoherent bremsstrahlung
radiation emitted in the amorphous target.
For larger emission cones the peaks become less pronounced as
well as the excess rate over the emission in the amorphous medium.

Qualitatively, these features can be understood in terms of the main
characteristics of the undulator radiation
(see, e.g.,  \cite{AlferovBashmakovCherenkov1989,Barbini_EtAl_1990}).
The radiation emitted by a particle from each period of an ideal
harmonic profile (\ref{Geometry:eq.07b}) interferes constructively
at some photon frequencies $\om_k$, which are integer multiples to the
frequency $\om_1$ of the first (fundamental) harmonic.
As a result, for each value of the emission angle $\theta$ the
spectral distribution consists of a set of equally spaced peaks
centered at $\om_k = k\om_1$ ($k=1,2,\dots$)
where the fundamental photon energy $\hbar\om_1$ can be
calculated as follows (see, e.g., \cite{Elleaume:RevSciInstrum_v63_p321_1991,CLS-book_2022}):
\begin{eqnarray}
\hbar\om_1(\theta)\, \mbox{[MeV]}
\approx
{ 9.5 \E^2 \, \mbox{[GeV]} \over \lambda_{\rm u}\, \mbox{[$\mu$m]}}
{1 \over 1 + K^2/2 + (\gamma\theta)^2 }
\label{MBN_Emission:eq.03}
\end{eqnarray}
Here, the undulator parameter $K$ is related to the average velocity of
the particle in the transverse direction.
In a CU this parameter accounts for both the undulator oscillations and
the channeling oscillations \cite{Dechan01}:
\begin{eqnarray}
K^2 = K_{\rm u}^2 + K_{\rm ch}^2
 \label{MBN_Emission:eq.02}
\end{eqnarray}
where $K_{\mathrm{u}}=2\pi \gamma a/\lamu$ and
$K_{\mathrm{ch}} \propto 2\pi \gamma  a_{\rm ch}/\lambda_{\mathrm{ch}}$.
For the positron channeling, assuming harmonicity of the channeling
oscillations, one can \textit{derive} the value of $K^2$  averaged over
the range $[0,d/2]$ of acceptable
values of the channeling amplitude:
$\langle K_{\rm ch}^2 \rangle  = {2\gamma U_0 / 3mc^2}$
(see, e.g., \cite{ChannelingBook2014}, Eq. (B.5)).
For the $U_0$ data from Table \ref{Si110.Table01} this relation produces
$\langle K_{\rm ch}^2  \rangle\approx 1$ for $\E=20$ GeV positrons in Si(110).

For small emission cones, $\gamma\theta_0 \ll 1$, the spectral
distribution $\d E(\theta_0)/\d(\hbar\om)$ is peaked in the close
vicinity to the maximum fundamental
photon energy $\hbar\om_1(0)$ which characterizes the on-axis emission,
i.e. $\theta=0$.
With increase in  $\theta_0$ more radiation is emitted with lower
energies so that the peaks in the spectral distribution become wider and
red-shifted.
This behaviour is seen in Figs. \ref{10f_Spectra.fig}  and
\ref{40f_Spectra.fig}.
In both figures the vertical dashed lines mark the position of
$\hbar\om_1(0)$, which was obtained from Eqs. (\ref{MBN_Emission:eq.03})
and (\ref{MBN_Emission:eq.02}).
To calculate the undulator parameter $K_{\mathrm{u}}$ for the
non-harmonic profiles
from Fig. \ref{Profiles_free_RC.fig} (top) the average values of the
bending amplitude were used: $\langle a \rangle \approx 20 d$ for
$\nu=10$ MHz and
$\langle a \rangle \approx 2 d$ for $\nu=40$ MHz.
The latter value corresponds to the amplitude averaged over the initial
3.5 mm segment of the
crystal since most of the positrons dechannel beyond this point.
It is seen that although Eqs. (\ref{MBN_Emission:eq.03}) and
(\ref{MBN_Emission:eq.02})
are written for the ideal profile (\ref{Geometry:eq.07b}), they
provide a good quantitative estimate for the peak position in case
of the acoustically excited periodic bending.

Another feature seen in Figs. \ref{10f_Spectra.fig}  and \ref{40f_Spectra.fig} concerns the relative decrease
in the enhancement of radiation emitted in the A-CU in comparison with
that in the amorphous medium as the emission cone increases.
Indeed, for both values of $\nu$ the factor,
$\d E_{\rm CU}/\d E_{\rm am}$,
calculated at the peak energies gradually decreases as $\theta_0$ becomes
larger.
The radiation formed in a segment of the trajectory is emitted
predominantly within the cone
$\theta \sim \gamma^{-1}$ along the vector of the instant velocity
$\bfv$.
Random multiple scattering from the atoms change the direction of the
velocity.
In an amorphous medium, the rms multiple scattering angle
$\langle \Theta \rangle$ can be estimated using
Eqs. (33.14-15) from Ref. \cite{ParticleDataGroup2018}.
For a 20 GeV positron in a 6 mm thick amorphous silicon the result is
$\langle \Theta \rangle \approx 150$ $\mu$rad, i.e. six times larger than
$\gamma^{-1}$.
Therefore, in this case only a small fraction of radiation is emitted
within the cones $\theta_0 \lesssim \gamma^{-1}$.
Positrons that experience planar channeling in an oriented crystal
move between two neighbouring crystal planes, i.e. in the spatial domain
where the atomic electron density is reduced.
As a result, the multiple scattering is suppressed
\cite{ScandaleEtAl:EPJC_v79_993_2019} and the change in the direction
of $\bfv$ is due to the channeling oscillations.
Hence, for these particles the natural emission cone must be
compared not with  $\langle \Theta \rangle$ but with Lindhard's
critical angle $\theta_{\rm L}$ which is approximately 50 $\mu$rad
for a 20 GeV positron
in a silicon (110) crystal, see Table \ref{Si110.Table01}.
As a result, the CU radiation from the channeling particles is emitted
in narrower cones along the direction of the incident beam.

\section{Conclusions \label{Conclusions}}

In this paper we have presented a computational study of the generation
of narrowband gamma-ray radiation emitted by ultra-relativistic
positrons channeling in an acoustically excited periodically bent
crystal.
We have proposed a novel A-CU device with tunable properties
based on the acousto-optic modulator scheme and have presented a complete
computational procedure for the design and characterisation of such a
device.
The A-CU under consideration is based on a silicon single crystal
in which a plane longitudinal traveling acoustic wave is generated by a
transducer along the [100] crystallographic direction.
The acoustic field causes a periodic deformation of the crystal lattice,
in particular a periodic bending of the (110) planes.
The amplitude and period  of the bending are controlled by the amplitude
and frequency of the acoustic pressure, respectively.
Ultra-relativistic positrons, incident on the crystal along the (110)
planar direction, can be accepted in the channeling regime and then
propagate through the crystal following the periodic profile of the
bent planes.
This motion results in the emission of narrowband gamma-ray radiation.

The simulated case studies refer to 20 GeV positrons propagating along
6.1 mm periodically bent Si(110).
The bending profiles were obtained by FEM simulations of the crystal
response to periodic acoustic excitation with 4 MPa amplitude and
10 MHz and 40 MHz frequencies.
The subsequent relativistic molecular dynamics simulations of the
positron propagation and the radiation emission showed that despite
the presence of non-harmonic components in the profiles there is strong
enhancement of the radiation at photon energies within narrow spectral
bands around 2 MeV and 10
MeV, for the two excitation frequencies, respectively.
The results clearly demonstrate the feasibility of the proposed A-CU
scheme and highlight the advantages of dynamic modulation, which is
tunable in terms of undulation amplitude and period.
In the near future, a real A-CU device will be implemented on the basis
of the presented methodology, which can be used in experiments for the
generation of gamma rays at various accelerator facilities.

\section{Acknowledgement}
The authors acknowledge financial support from the European
Commission's Horizon Europe-EIC-Pathfinder-Open TECHNO-CLS (G.A.
101046458) project and from the H2020 RISE-NLIGHT project (G.A.
872196).
The possibility of performing computer simulations at the Goethe-HLR
cluster of the Frankfurt Center for Scientific Computing is gratefully
acknowledged.
The HMU team acknowledge the support with computational time granted by the Greek Research and Innovation Network (GRNET) in the National HPC facility ARIS under project ID pr016025-LaMPIOS III.

\vspace*{0.3cm}
\noindent
\textbf{Competing Interests.}
The authors do not declare any conflicts of interest, and there is no
financial interest to report.

\vspace*{0.3cm}
\noindent
\textbf{Author Contribution.}
\\
\textbf{KK:} Development of the AOM-type A-CU scheme;
Design of the experimental device and setup;
Calculations of the device specifications;
Analysis and processing of the FEM data;
Writing initial draft and further editing.
\\
\textbf{EK:} FEM simulations; Analysis of the FEM data.
\\
\textbf{VD:} Supervision of the FEM simulations;
Design of the A-CU device; Analysis of the FEM data.
\\
\textbf{EKK:} Calculations
of the device specifications; Arrangement of the initial draft.
\\
\textbf{MB:} Development of the A-CU scheme;
Design of the experimental device.
\\
\textbf{NAP:} Conceived the concept of the
AOM-type A-CU; Overview and supervision of the work;
Development of the A-CU scheme and
the design of the experimental device and setup.
\\
\textbf{MT:} Analysis of the FEM
data; Design of the experimental device and setup.
\\
\textbf{GBS:} MBN Explorer Software;
Rel-MD Methodology; Simulation Algorithms.
\\
\textbf{AVK:}
Conceptualization and feasibility analysis of
the A-CU scheme;
Rel-MD methodology, simulations and analysis;
Writing – review \& editing.
\\
\textbf{AVS:} Project administration;
Conceptualization and feasibility analysis of
the A-CU scheme;
Rel-MD methodology, simulations and analysis;
Writing – review \& editing.

All authors reviewed the final manuscript.


\section*{References}

\bibliography{references}
\end{document}